\documentclass[aps,prd,reprint,showpacs,showkeys,preprintnumbers,superscriptaddress]{revtex4-2}
\usepackage{graphicx,epsfig,color,xcolor}
\usepackage[english]{babel}
\usepackage{amssymb,amsfonts,amsmath,physics}
\usepackage{mathtools}
\usepackage{siunitx}
\usepackage{verbatim}
\usepackage[normalem]{ulem}
\usepackage{soul}
\usepackage{lipsum, babel}
\usepackage{acronym}
\usepackage{aas_macros}

\usepackage[colorlinks=true
,urlcolor=DARKBLUE
,anchorcolor=DARKBLUE
,citecolor=DARKBLUE
,filecolor=DARKBLUE
,linkcolor=DARKBLUE
,menucolor=DARKBLUE
,pagecolor=DARKBLUE
,linktocpage=true
,pdfproducer=medialab
,pdfa=true
,bookmarks=true
]{hyperref}

\DeclareMathOperator{\erfc}{erfc}
\DeclareMathOperator{\argmin}{argmin}

\newcommand{\be}{\begin{equation}} \newcommand{\ee}{\end{equation}}
\newcommand{\bea}{\begin{eqnarray}} \newcommand{\eea}{\end{eqnarray}}

\newcommand{\cs}{c_{\mathrm{s}}}
\newcommand{\csmin}{c_{\mathrm{s},\mathrm{min}}}

\newcommand{\pk}{\mathrm{pk}}

\newcommand{\PBH}{\mathrm{PBH}}

\newcommand{\tot}{\mathrm{tot}}
\newcommand{\eq}{\mathrm{eq}}
\newcommand{\sh}{\mathrm{sh}}
\newcommand{\rad}{\mathrm{rad}}
\newcommand{\umax}{\mathrm{max}}
\newcommand{\uapprox}{\mathrm{approx}}
\newcommand{\GW}{\mathrm{GW}}
\newcommand{\EW}{\mathrm{EW}}
\newcommand{\cut}{\mathrm{cut}}

\newcommand{\calA}{\mathcal{A}}
\newcommand{\ub}{\mathrm{b}}

\newcommand{\uc}{\mathrm{c}}

\newcommand{\ue}{\mathrm{e}}

\newcommand{\uf}{\mathrm{f}}

\newcommand{\calH}{\mathcal{H}}

\newcommand{\ui}{\mathrm{i}}

\newcommand{\um}{\mathrm{m}}

\newcommand{\calN}{\mathcal{N}}

\newcommand{\calP}{\mathcal{P}}

\newcommand{\ur}{\mathrm{r}}
\newcommand{\ut}{\mathrm{t}}

\newcommand{\uW}{\mathrm{W}}

\newcommand{\bae}[1]{\begin{align} #1 \end{align}}

\newcommand{\beae}[1]{\begin{equation}\begin{aligned} #1 \end{aligned}\end{equation}}

\newcommand{\bme}[1]{\begin{multline} #1 \end{multline}}

\newcommand{\bmbe}[1]{\begin{multlined}[b] #1 \end{multlined}}

\definecolor{MONZA}{HTML}{CF000F}
\definecolor{DARKBLUE}{HTML}{00008b}
\definecolor{DARKMAGENTA}{HTML}{8b008b}
\definecolor{DARKCYAN}{HTML}{008B8B}
\definecolor{DARKORANGE}{HTML}{FF8C00}
\definecolor{OBSERVATORY}{HTML}{049372}
\definecolor{GREENBAMBOO}{HTML}{006442}
\definecolor{TURQUOISE}{HTML}{36D7B7}
\definecolor{JUNGLEGREEN}{HTML}{26C281}

\begin{document}

%%%%%%%%%%%%%%%%%%%%%%%%%%%%%%%%%%%%%%%%%%%%%%%%%%%
%%%%%%%%%%%%%%%%%%%%%%%%%%%%%%%%%%%%%%%%%%%%%%%%%%%
\title{Primordial Black Holes and Induced Gravitational Waves from a Smooth Crossover beyond Standard Model}

\author{Albert Escrivà}
\email{escriva.manas.albert.y0@a.mail.nagoya-u.ac.jp}
\affiliation{\mbox{Division of Particle and Astrophysical Science, Graduate School of Science,} \\ Nagoya University, Nagoya 464-8602, Japan}
\author{Yuichiro Tada}
\email{tada.yuichiro.y8@f.mail.nagoya-u.ac.jp}
\affiliation{Institute for Advanced Research, Nagoya University, \\
Furo-cho Chikusa-ku, 
Nagoya 464-8601, Japan}
\affiliation{\mbox{Division of Particle and Astrophysical Science, Graduate School of Science,} \\ Nagoya University, Nagoya 464-8602, Japan}
\author{Chul-Moon Yoo}
\email{yoo.chulmoon.k6@f.mail.nagoya-u.ac.jp}
\affiliation{\mbox{Division of Particle and Astrophysical Science, Graduate School of Science,} \\ Nagoya University, Nagoya 464-8602, Japan}
%%%%%%%%%%%%%%%%%%%%%%%%%%%%%%%%%%%%%%%%%%%%%%%%%%%
%%%%%%%%%%%%%%%%%%%%%%%%%%%%%%%%%%%%%%%%%%%%%%%%%%%
\begin{abstract}
\Acp{GW} induced by primordial fluctuations can be affected by the modification of the sound speed $\cs^2$ and the equation of state parameter $w$ once the curvature fluctuations reenter the cosmological horizon. That softening can also significantly boost the production of \acp{PBH} at the mass scale where the softening arises. In this work, we consider a hypothetical softening of $w$ and $\cs^2$ caused by a smooth crossover beyond Standard Model theories, for what we numerically compute the secondary induced \acs{GW} considering the case of a flat scale-invariant power spectrum. We find that if the amplitude of the power spectrum is sufficiently large, the characteristic feature of the \ac{GW} signal caused by the smooth crossover can be detected by future space-based gravitational wave interferometers and differentiated from the pure radiation case. At the same time, depending on the mass scale where the crossover occurs, such a scenario can have compatibility with \acp{PBH} being all the dark matter when $\mathcal{A} \sim \mathcal{O}(10^{-3})$, with a mass function very sharply peaked around the horizon mass scale of the minimum of the sound speed. Our results show that the \ac{GW} signal can be used to resolve the existing degeneracy of sharply peaked mass function caused by peaked power spectrums and broad ones in the presence of softenings of $w$ and $\cs^2$.
\end{abstract}
\keywords{Induced Gravitational Waves, Primordial Black Holes, Dark Matter}
\pacs{%26.60.Kp, 21.65.Qr, 11.25.Tq 
}

\maketitle
\acresetall

\acrodef{GW}{gravitational wave}
\acrodef{PT}{phase transition}
\acrodef{SC}{smooth crossover}
\acrodef{SM}{Standard Model}
\acrodef{QCD}{Quantum Chromodynamics}
\acrodef{EW}{electroweak}
\acrodef{CMB}{cosmic microwave background}
\acrodef{PBH}{primordial black hole}
\acrodef{DM}{Dark Matter}
\acrodef{FLRW}{Friedmann‐-Lema\^itre--Robertson--Walker}
\acrodef{MS}{Misner--Sharp}

%%%%%%%%%%%%%%%%%%%%%%%%%%%%%%%%%%%%%%%%%%%%%%%%%%%
%%%%%%%%%%%%%%%%%%%%%%%%%%%%%%%%%%%%%%%%%%%%%%%%%%%
\section{Introduction}
%%%%%%%%%%%%%%%%%%%%%%%%%%%%%%%%%%%%%%%%%%%%%%%%%%%
Since the first \ac{GW} detection~\cite{LIGOScientific:2016aoc}, \acp{PT} in the very early Universe have obtained increasing attention due to the possibility of testing them with planned future space-based \acp{GW} interferometers~\cite{LISA:2017pwj,Seto:2001qf,Yagi:2011wg}. Several studies focus on first-order \acp{PT}, during which bubbles are nucleated, and its collapse, collision, and expansion can lead to a detectable stochastic background of \acp{GW}~\cite{Kosowsky:1992vn,Kamionkowski:1993fg,Caprini:2009fx,Hindmarsh:2013xza,Huang:2016odd,Hindmarsh:2020hop,Guo:2020grp,Kalogera:2021bya,Caprini:2019egz,Cai:2017cbj,Athron:2023xlk}. Another interesting possibility is focusing on \ac{SC}, not a \ac{PT}. For instance, within the \ac{SM}, the deconfinement in \ac{QCD}~\cite{Aoki:2006we} and \ac{EW} phase transitions~\cite{Kajantie:1996mn, Laine:1998vn, Rummukainen:1998as} are actually \ac{SC} \footnote{See Refs.~\cite{Carena:1996wj,Laine:1998qk,Grojean:2004xa,Huber:2006ma,Profumo:2007wc,Laine:2012jy,Damgaard:2015con} for realizations where the \ac{EW} becomes first order.}.

Induced \acp{GW}, associated with primordial fluctuations, can be a probe of the existence of primordial scalar curvature fluctuations at much smaller scales than the \ac{CMB} scale~\cite{Assadullahi:2009jc,2011PhRvD..83h3521B,Inomata:2019ivs,Pi:2020otn} (see Ref.~\cite{Domenech:2021ztg} for a review). Their spectrum not only depends on the specific shape of the power spectrum that generates the curvature fluctuation but also on the thermal history~\cite{Baumann:2007zm,Tomikawa:2019tvi,Ota:2020vfn,Domenech:2020ers,Kohri:2018awv,Inomata:2019zqy,Inomata:2019ivs,Hajkarim:2019nbx,Cai:2019cdl,Domenech:2019quo,2020JCAP...08..017D,Dalianis:2020cla,Pearce:2023kxp}. Since the induced \acp{GW} are affected by the modification of the equation-of-state parameter $w$ and sound speed $\cs$ (see Refs.~\cite{Hajkarim:2019nbx,Abe:2020sqb,Abe:2023yrw} for a numerical study focusing on the \ac{QCD} crossover), it can be a direct prove of the existence of a crossover beyond \ac{SM} that modifies $w$ and $\cs^2$ in a specific time-scale during the early Universe. 

Additionally, induced \acp{GW} can be an indirect probe of the existence of \acp{PBH}~\cite{Saito:2008jc,2010PhRvD..81b3517B,2010PThPh.123..867S,2011PhRvD..83h3521B,Alabidi:2012ex,Inomata:2016rbd,Bartolo:2018evs,Bhattacharya:2019bvk,2019PhRvL.122t1101C}, formed in the very early Universe~\cite{1975ApJ...201....1C} (see Ref.~\cite{Escriva:2022duf} for a review). For sufficiently large curvature fluctuations, \acp{PBH} can constitute all the \ac{DM} or a significant fraction of it~\cite{Chapline:1975ojl}. Since their abundance is exponentially sensitive to the threshold for a perturbation to collapse, a threshold reduction enhances \ac{PBH} production. This results in a sharply-peaked mass function at the scale where a \ac{SC} occurs, as shown in Ref.~\cite{Escriva:2022yaf} (see also Refs.~\cite{Jedamzik:1996mr,Byrnes:2018clq,Carr:2019kxo,Juan:2022mir,Escriva:2022bwe,Franciolini:2022tfm} related to the \ac{QCD} crossover and Ref.~\cite{Lu:2022yuc} for a beyond the SM extension).

%In this work, we focus on the possibility of finding a potential gravitational wave signature associated with a hypothetical \ac{SC} beyond \ac{SM} for temperatures above $T \gtrsim \SI{0.2}{TeV}$ (corresponding to horizon mass $M_H \lesssim 10^{-6} M_{\odot}$), through the numerical computation of the spectrum of the scalar-induced \acp{GW} affected by that \ac{SC}, and find its implications in the \ac{PBH} scenario. The detection of a \ac{GW} signature caused by the \ac{SC} beyond \ac{SM} would imply not only the existence of new physics that is sourcing the modification of the thermal history from a pure radiation epoch but also a possible indirect proof of a large production of \acp{PBH} in the mass range where the crossover occurs if the curvature fluctuation is sufficiently large. This is the motivation of our study: finding a potential observation signature of the \ac{SC} that could be observed in future \ac{GW} interferometers with the corresponding prediction of the \ac{PBH} mass function. In particular, we focus our computation on the \ac{GW} signature induced from a flat scale-invariant power spectrum, whose modification of the \ac{GW} signal can be only associated with the change in the thermal history and not from the specific shape of the power spectrum since the spectrum of the \acp{GW} is independent on the wavenumber scale $k$. We adopt the Planck unit $c=\hbar=G=1$ throughout the paper.

This work explores the potential gravitational wave signature of a hypothetical \ac{SC} beyond \ac{SM} for temperatures $T \gtrsim \SI{0.2}{TeV}$ (corresponding to horizon mass $M_H \lesssim 10^{-6} M_{\odot}$). We numerically compute the spectrum of scalar-induced \acp{GW} affected by this \ac{SC} and examine its implications in the \ac{PBH} scenario. Detecting a \ac{GW} signature from the \ac{SC} would imply new physics modifying the thermal history from a pure radiation epoch and could indirectly support a large \ac{PBH} production in the mass range of the crossover if the curvature fluctuation is sufficiently large. Our motivation is to identify a potential observational signature of the \ac{SC} that may be observed in future \ac{GW} interferometers, along with predicting the corresponding \ac{PBH} mass function. Specifically, we focus on the \ac{GW} signature induced by a flat scale-invariant power spectrum, where modifications arise solely from changes in the thermal history, not the power spectrum's specific shape. Throughout the paper, we use Planck units with $c=\hbar=G=1$.

%%%%%%%%%%%%%%%%%%%%%%%%%%%%%%%%%%%%%%%%%%%%%%%%%%%
%%%%%%%%%%%%%%%%%%%%%%%%%%%%%%%%%%%%%%%%%%%%%%%%%%%
\section{The model}
\begin{figure}[t] 
    \centering
    \includegraphics[width=0.4\textwidth]{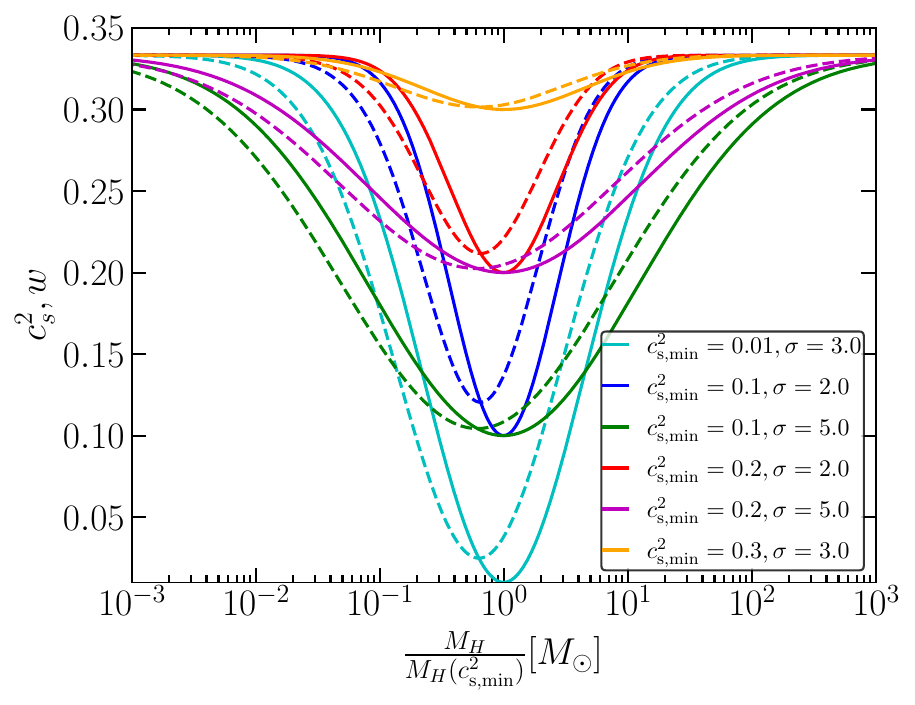}
    \caption{Template of $\cs^2$ (solid line) and $w$ (dashed line) for different parameters. The horizontal axis is labeled by the horizon mass normalized by its value at the minimum of $\cs^2$.}
    \label{fig:eq_of_state}
\end{figure}

%%%%%%%%%%%%%%%%%%%%%%%%%%%%%%%%%%%%%%%%%%%%%%%%%%%
%%%%%%%%%%%%%%%%%%%%%%%%%%%%%%%%%%%%%%%%%%%%%%%%%%%
To modulate the hypothetical \ac{SC} beyond \ac{SM}, we consider that the softening of the sound speed $\cs^2=\pdv*{p}{\rho}$ ($p$ is the pressure and $\rho$ is the energy density of the Universe) follows a log-normal template as a function of $\rho$: 
\begin{equation}
    \cs^2(\tilde{\rho}) = w_0- (w_0-\csmin^2) \exp\left[{\frac{-(\ln \tilde{\rho})^2}{2 \sigma^2} }\right],
\label{eq:cs}
\end{equation}
with $\tilde{\rho}=\rho/\rho_\uc$, where the model parameters $\rho_\uc$, $\csmin^2$, and $\sigma$ are the location of the minimum of $\cs^2$, its value, and the width of the \ac{SC}, respectively. The template of Eq.~\eqref{eq:cs} is mainly motivated by crossovers built from holographic models \cite{Escriva:2022yaf} beyond \ac{SM} theories. It also has similarities to the case of QCD crossover within \ac{SM}~\cite{Aoki:2006we} and %beyond a 
its beyond-\ac{SM} extension~\cite{Lu:2022yuc}. The convenient choice of the parameters allows several realizations that can approximately fit the different models. Then, the corresponding equation of state $w(\rho)=p/\rho$ can be obtained by integrating Eq.~\eqref{eq:cs},
\begin{equation}
    w(\tilde{\rho}) =w_0-\frac{\sigma}{\tilde{\rho}} \sqrt{\frac{\pi}{2}}\ue^{\sigma^2/2}(w_0-c^2_{\rm s,min}) \erfc\left[ \frac{\sigma^2-\ln(\tilde{\rho})}{\sqrt{2}\sigma} \right],
\label{eq:w}
\end{equation}
with $w_0=1/3$ during the radiation-dominated era. Different realizations of Eqs.~\eqref{eq:cs} and \eqref{eq:w} can be found in Fig.~\ref{fig:eq_of_state}.
Fixing $w(\rho)$, we can obtain the \ac{FLRW} background dynamics by solving
\beae{
    \rho'(\eta) &= -\sqrt{24 \pi} (1+w(\rho))a(\eta) \rho^{3/2}(\eta), 
    \\
    a'(\eta) &= \sqrt{\frac{8 \pi \rho(\eta)}{3}}a^2(\eta),
    \label{eq:background_dynamics}
}
where $\eta$ is the conformal time $\eta = \int a^{-1} \dd{t}$ and the Hubble factor in terms of the conformal time is defined as $\mathcal{H} = a'/a $, where $'$ denotes derivative $\partial_{\eta}$. Eq.~\eqref{eq:background_dynamics} needs to be solved numerically taking into account Eq.~\eqref{eq:w}. We consider a source for the primordial Gaussian curvature fluctuation $\zeta$, given by a flat-scale invariant power spectrum extending over scales much shorter than the \ac{CMB} scale,
\begin{equation}
    \mathcal{P}_{\zeta}(k) = \mathcal{A}.
    \label{eq:ps}
\end{equation}
%
%%%%%%%%%%%%%%%%%%%%%%%%%%%%%%%%%%%%%%%%%%%%%%%%%%%
\section{Primordial Black Hole production }
%%%%%%%%%%%%%%%%%%%%%%%%%%%%%%%%%%%%%%%%%%%%%%%%%%%
%%%%%%%%%%%%%%%%%%%%%%%%%%%%%%%%%%%%%%%%%%%%%%%%%%%
\label{sec:pbh}
The softening of $w$ and $\cs^2$ affects the \ac{PBH} production by the horizon reentry of a large superhorizon adiabatic curvature fluctuation $\zeta$ (which remains frozen on a superhorizon scale) generated in the early Universe. In this scenario, \ac{PBH} formation should be associated with a rare, high peak (in position space) of $\zeta$ so that \acp{PBH} remain well subdominant during the radiation-dominated era. If one supposes $\zeta$ to follow the Gaussian statistics, such a high peak typically forms a spherically symmetric shape, which can be locally modeled by the spacetime~\cite{Shibata:1999zs}
\begin{equation}
    \dd{s^2} = -\dd{t^2} + a^{2}(t) \ue^{2 \zeta(r)}\left[ \dd{r^2}+ r^2 \left( \dd{\theta^2}+\sin^2(\theta) \dd{\phi^2}  \right)  \right],
    \label{eq:metric_super_horizon}
\end{equation}
where $r=0$ is placed at the local maximum of $\zeta$. The typical profile and probability of $\zeta(r)$ should also be characterized only by the power spectrum~\eqref{eq:ps} in the Gaussian case, which is known as the peak theory~\cite{1986ApJ...304...15B}. We follow the approach of Ref.~\cite{Yoo:2020dkz}, which applies peak theory~\cite{1986ApJ...304...15B} to account for peaks on 
$\Delta\zeta$ rather than $\zeta$~\cite{pic_original} to 
drop the contamination by physically irrelevant long-wavelength mode.
The cloud-in-cloud effect can also be treated by implementing a top-hat window function $W(k) = \Theta(k_\uW-k)$ with a UV cut-off scale $k_\uW$. Specifically, we eliminate irrelevant contributions from smaller-scale perturbations at a PBH mass scale of interest using the window function. Eventually, the true  
\ac{PBH} mass function $f_\PBH$ is obtained as the envelope curve for various $k_\uW$:
\bae{
\label{eq:mass_function_eq}
	f_\PBH(M)=\max_{k_\uW}\Bqty{f_\PBH(M\mid k_\uW)}.
}
$f_\PBH(M)\dd{\ln M}$ is the current \ac{PBH} abundance in the mass range $[M,M\ue^{\dd{\ln M}}]$ normalized by the total \ac{DM} abundance. 

Given $k_\uW$, the \textit{typical profile} of the curvature fluctuation reads $\zeta(r) = \mu_2 g(x;k_{\bullet})$, where $\mu_2 = -(2/ k^2_\uW) \eval{\Delta\zeta}_{r=0}$ (the normalized peak value of $\Delta\zeta$) and $k_\bullet=-\eval{\Delta\Delta\zeta}_{r=0}/(\eval{\Delta\zeta}_{r=0})$ (the curvature scale of the peak) are both random variables and $x=r/r_\um$ is the normalized radius by $r_\um$, the location of the peak value of the compaction function during radiation domination at super-horizon scales, $\mathcal{C}(x %=1
)=%(2/3)
\frac{2}{3}\qty(1-\qty(1+ \mu_{2} %\eval{
\partial_{x} g(x;\kappa)%}_{x=1} 
)^2)$~\cite{Shibata:1999zs,2015PhRvD..91h4057H} where $\kappa=k_\bullet/k_\uW$. In the scale-invariant case~\eqref{eq:ps}, the template $g$ is given by
\begin{multline} 
    \label{eq:typical_profile_zeta}
    g(x;\kappa) = \frac{6 }{x^4 \lambda^4}\Biggl( -12 \kappa^2 + x^2 \lambda^2 (3-4 \kappa^2) + \\
    4 \left(3 \kappa^2-2\right)  x \lambda \sin \left(x \lambda \right)+8  + \\ 
    \left[12 \kappa^2 -8 + x^2 \lambda^2 (1-2 \kappa^2)\right] \cos \left(x \lambda \right) \Biggr),
\end{multline}
with $\lambda = k_\uW r_\um$. %, and $\kappa = k_{\bullet}/k_\uW$. 
The (normalized) maximum radius $\lambda$ is found by solving the maximum compaction condition, $\left[ \partial_x g(x;\kappa)+\partial^2_x g(x;\kappa) \right]_{x=1}= 0$, for each $\kappa$.

\begin{figure} 
    \centering
    \includegraphics[width=0.4\textwidth]{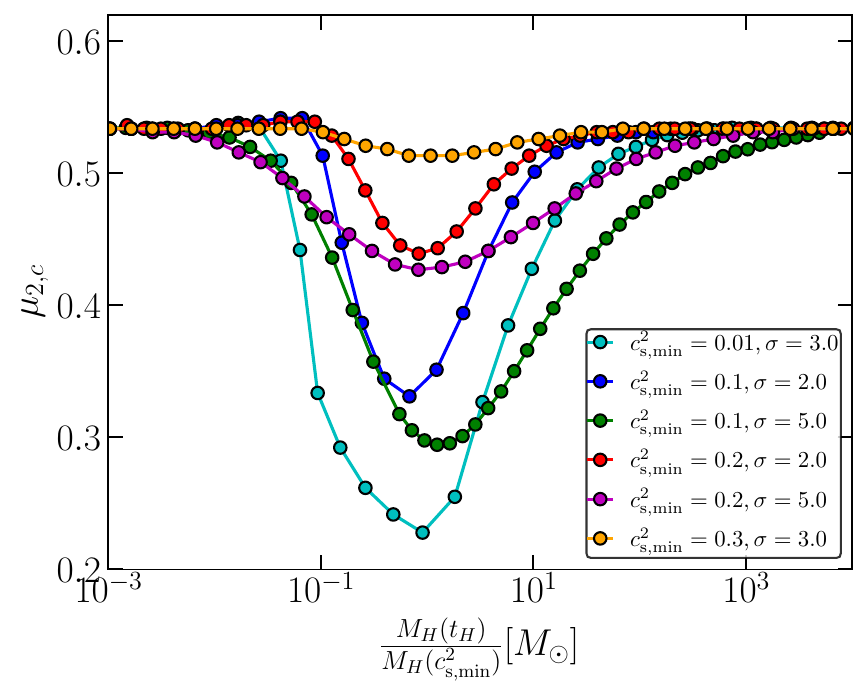}
    \caption{Threshold values $\mu_{2,\uc}$ for the curvature profile of Eq.~\eqref{eq:typical_profile_zeta} and for the different templates of the \ac{SC}.}
    \label{fig:thresholds}
\end{figure}

For specific configurations of Eq.~\eqref{eq:typical_profile_zeta}, we make numerical simulations following Refs.~\cite{Escriva:2019nsa,Escriva:2022yaf,Escriva:2022bwe} to obtain the threshold values $\mu_{2,\uc}$ on the curvature peak $\mu_2$ to form a \ac{PBH}, taking into account the reduction in pressure gradients from the \ac{SC} (see the supplemental material for details). 
The results are shown in Fig.~\ref{fig:thresholds}, where we fixed $\kappa=\kappa_\ut\approx0.707$ with $\lambda \approx 4.16$ for which \acp{PBH} are most likely produced (numerically found).
The overall scale $k_\uW$ is labeled by the horizon mass $M_H$ at the horizon reentry $t_H$. One can observe that the reduction in the threshold values is increased when the duration of the \ac{SC} is more extended and/or the softening of $w$ and $\cs^2$ is larger. %\textcolor{red}{Notice that the reduction is much larger than the one found from the \ac{QCD} crossover~\cite{Escriva:2022bwe}.}

The \ac{PBH} formation rate for $\kappa_\ut$ can be approximated by~\cite{Yoo:2020dkz}
\begin{multline} 
    \beta_{0,\umax}^{\uapprox}:=\Biggl[\frac{ \kappa \lambda^3 }{36 \sqrt{\pi}(6 \kappa^4-8 \kappa^2+3)}\ue^{3\mu_{2} g(1;\kappa) } \\
    \times  f \left( \sqrt{\frac{2}{\mathcal{A}}} \kappa^2 \mu_2 \right) P_1\left( \frac{\mu_{2}}{\sqrt{A}}, \sqrt{\frac{2}{\mathcal{A}}}\kappa^2\mu_{2}\right) \\
    \times \abs{\dv{\kappa}\ln \lambda+\mu_2 \dv{\kappa}g_\um}^{-1}
    \Biggr]_{\kappa=\kappa_\ut,~\mu_2=\mu_{2,\uc}(k_\ut)}.
    \label{eq:beta_max_aprox_main}
\end{multline}

The correlated Gaussian $P_1$ and function $f$ are given in the supplemental material. The formation rate is related to the abundance by $f_\PBH(M_\ut\mid k_\uW)=\sqrt{M_\ut/M_{\eq}}\beta_{0,\umax}^{\uapprox}$, where the mass $M_\ut$ corresponding to $\kappa_\ut$ is given by $M_\ut = M_\eq k^2_\eq \lambda^2(\kappa_\ut)\ue^{2 \mu_{2,\uc} g(1;\kappa_\ut)}/k^2_\uW$ with the mode of horizon reentry at the equality, $k_\eq \approx \SI{0.01}{Mpc^{-1}}$, and the corresponding horizon mass, $M_\eq \approx \num{2.8e17} M_{\odot}$~\cite{Planck:2018vyg}.
Notice that Eq.~\eqref{eq:beta_max_aprox_main} shows no explicit dependence on the scale $k_\uW$ thanks to the flat spectrum assumption.  
The non-trivial scale dependence only comes from the effective reduction of the threshold $\mu_{2,\uc}$ by \ac{SC}.

\begin{figure}
    \centering
    \includegraphics[width=0.40\textwidth]{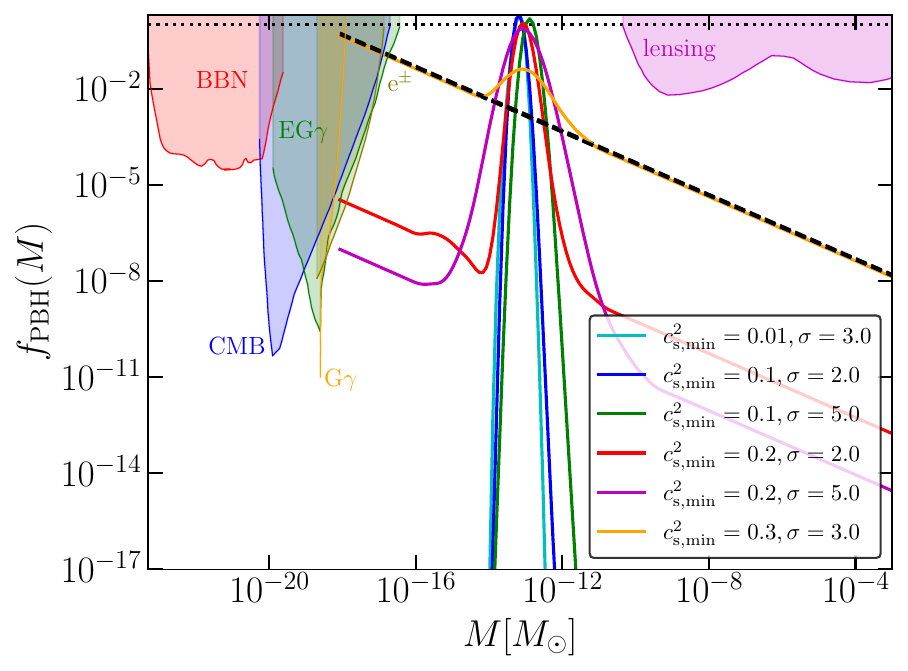}
    \includegraphics[width=0.40\textwidth]{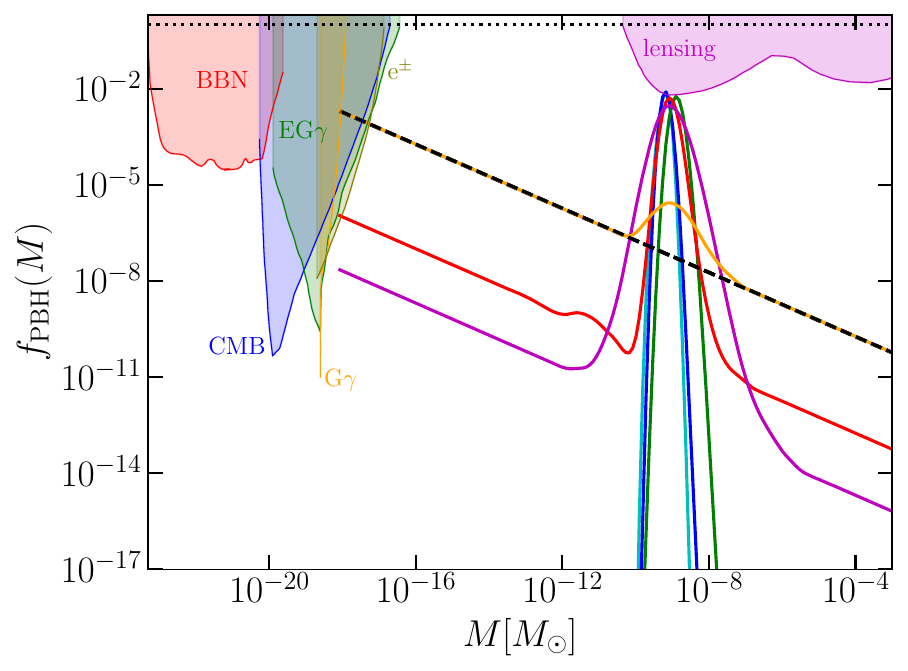}
    \caption{\ac{PBH} mass functions for models \textbf{A} ($\rho_\uc^{1/4}\approx\SI{1.7e6}{GeV}$; top-panel) and \textbf{B} ($\rho_\uc^{1/4}\approx\SI{17}{TeV}$; bottom-panel), considering the \ac{SC} templates of Fig.~\ref{fig:eq_of_state}. The different coloured regions correspond to constraints taken from Ref.~\cite{2021RPPh...84k6902C}. The black-dashed lines correspond to the pure radiation case $w=\csmin^2=1/3$.}
    \label{fig:PBH_mass_function}
\end{figure}

\begin{table} 
    \begin{tabular}{|c|c|c|c|}
        \hline
        $c^2_{s,\min}$ & $\sigma$ & $\mathcal{A}^{\textbf{A}}/10^{-3}$ & $\mathcal{A}^{\textbf{B}}/10^{-3}$ \\ \hline
        0.01 & 3.0 &  1.045 & 1.021 \\ \hline
        0.1 & 2.0 &  2.136 & 2.083 \\ \hline
        0.1 & 5.0 & 1.672 & 1.629 \\ \hline
        0.2 & 2.0 & 3.668 & 3.596 \\ \hline
        0.2 & 5.0 & 3.420 & 3.339 \\ \hline
        0.3 & 3.0 & 4.582 & 4.090 \\ \hline
    \end{tabular}
    \caption{The parameters of the amplitude of the power spectrum used in our models \textbf{A} (namely $\mathcal{A}^{\textbf{A}}$) and model \textbf{B} (namely $\mathcal{A}^{\textbf{B}}$). As a reference, the values for a pure radiation-domination era are $\mathcal{A}^{\textbf{A}}_{\rad} \approx  \num{4.58e-3}$ and $\mathcal{A}^{\textbf{B}}_{\rad} \approx \num{4.09e-3}$.}
    \label{table:parametters}
\end{table}

Varying the smoothing scale $k_\uW$, one can draw the mass function $f_\PBH(M)$. Example results 
are shown in Fig.~\ref{fig:PBH_mass_function}. We have considered two models, Model \textbf{A}: $M_H(\csmin^2) = 10^{-13}M_{\odot}$ ($\rho_\uc^{1/4}\approx\SI{1.7e6}{GeV}$) and Model \textbf{B}: $M_H(\csmin^2) = 10^{-9}M_{\odot}$ ($\rho_\uc^{1/4}\approx\SI{17}{TeV}$). The amplitude $\calA$ of the power spectrum is fixed so that $f_\PBH^\tot=1$ in the model \textbf{A} and $f_\PBH^\tot=\num{4e-3}$ to avoid the observational constraints from lensing in the model \textbf{B}, whose explicit values for each parameter set are shown in Table~\ref{table:parametters}. Notice that we have put a global UV cut-off $k_{\rm cut} \approx 10^{18} k_\eq$ (in terms of horizon mass $M_{H}(k_{\rm cut}) \approx \num{1.2e-19}M_{\odot}$) in the power spectrum to avoid the existing observational constraints.

The figure shows that the reduction on the thresholds $\mu_{2,\uc}\qty(M_H(t_H))$ significantly enhances the production of \acp{PBH} at the mass scale where the transition occurs (as also found in Ref.~\cite{Escriva:2022yaf}), inducing a significant sharp peak in the mass function. Furthermore, precisely when the reduction on $\mu_{2,\uc}$ is sufficiently significant, the mass function can be extremely sharp as if the primordial power spectrum has a strong peak and avoids all the low PBH mass range constraints for the same fixed $f_{\PBH}^{\rm tot}$ \footnote{Practically, the constraints can also be avoided by lowering the UV cut-off $k_\cut$.}. For $c^2_{\rm s,min} \approx 0.1$ we can already avoid the constraints even without introducing $k_\cut$.

\section{Scalar-induced gravitational wave}
\label{sec:computation_gws}
The strong \ac{SC} leaves imprints also in the scalar-induced \acp{GW}. We follow the computation approach of Refs.~\cite{Abe:2020sqb,Abe:2023yrw} that we briefly review here. The current (denoted by the time $\eta_{0}$) energy density of the induced GWs is given by
\bme{
        \Omega_{\rm GW}(k, \eta_0)h^2 \\ 
        =\Omega_{\ur,0}h^2\left(\frac{a_{ \rm sh} \mathcal{H}_\sh}{a_\uf \mathcal{H}_\uf}\right)^2\frac{1}{24} \left(\frac{k}{\mathcal{H}_\sh}\right)^2 \overline{\mathcal{P}_h(k,\eta_\sh)}.
     \label{eq:omega_gw_k_eta0}
}
where $\Omega_{\ur,0}h^2 \approx \num{4.2e-5}$ since $\Omega_{\ur,0}$ is the current radiation energy density parameter and $h$ is the renormalized Hubble parameter $h = H_0 / (\SI{100}{km.s^{-1}.Mpc^{-1}})$. The subscript ``$\sh$" denotes the time once the mode $k$ is well after the horizon crossing time, and ``$\uf$" denotes the time when we recover the pure radiation period after the \ac{SC}. 
$\overline{\mathcal{P}_h(k,\eta)}$ is the time averaged GW power spectrum $\calP_h(k,\eta)$ per period around $\eta$. The power spectrum $\calP_h(k,\eta)$ of the induced \acp{GW} is computed as
\bme{\label{eq: ps_vu}
    \calP_h(k,\eta)=\frac{64}{81a^2(\eta)}\int_{|k_1-k_2|\leq k\leq k_1+k_2}\hspace{-50pt}\dd{\ln k_1}\dd{\ln k_2}I^2(k,k_1,k_2,\eta) \\
    \times\frac{\left(k_1^2-(k^2-k_2^2+k_1^2)^2/(4k^2)\right)^2}{k_1k_2k^2}\calP_\zeta(k_1)\calP_\zeta(k_2),
}
where $I(k,k_1,k_2,\eta)$ is the kernel function defined as 
\bae{\label{eq: kernel_vu}
	&I(k,k_1,k_2,\eta)\!=\!k^2\int_0^\eta\!\dd{\tilde{\eta}} \, a(\tilde{\eta})G_k(\eta,\tilde{\eta})\biggl[2\Phi_{k_1}(\tilde{\eta})\Phi_{k_2}(\tilde{\eta}) \nonumber \\
    &\quad\left.+\frac{4}{3(1+w(\tilde{\eta}))}\left(\!\Phi_{k_1}(\tilde{\eta})\!+\!\frac{\Phi_{k_1}^\prime(\tilde{\eta})}{\calH(\tilde{\eta})}\!\right)\left(\!\Phi_{k_2}(\tilde{\eta})\!+\!\frac{\Phi_{k_2}^\prime(\tilde{\eta})}{\calH(\tilde{\eta})}\!\right)\!\right].
}
with the tensor Green's function $G_k(\eta,\eta')$ and the scalar transfer function $\Phi_k(\eta)$.

In a practical computation, Green's function $G_{k}(\eta,\eta')$ can be obtained by the combination of two independent homogeneous solutions $g_{1k}$ and $g_{2k}$ as
\begin{equation}
\label{eq:method_gf}
    G_k(\eta, \Tilde{\eta}) =
    \frac{1}{\calN_k}\left[g_{1k}(\eta) g_{2k}(\Tilde{\eta})-g_{1k}(\Tilde{\eta})g_{2k}(\eta)\right]\Theta(\eta-\tilde{\eta}),
\end{equation}
where $\calN_k$ is a constant $\mathcal{N}_k = g'_{1k}(\tilde{\eta})g_{2k}(\tilde{\eta})-g_{1k}(\tilde{\eta})g'_{2k}(\tilde{\eta})$ and $g_{jk}$ are obtained by solving \footnote{Green's function~\eqref{eq:method_gf} does not depend on the choice of the initial condition for $g_{jk}$.}
\begin{equation}
	\pqty{\partial^{2}_{\eta} +k^2-\frac{1-3w(\eta)}{2} \mathcal{H}^2(\eta)}g_{jk}(\eta)=0.
    \label{eq:lambda_operator}
\end{equation}
During the pure radiation era, we can make the following choices $g_{1k}=\sin(k \eta)$ and $g_{2k}= \cos(k \eta)$ as two independent solutions.

On the other hand, the scalar transfer function $\Phi_{k}(\eta)$ is a solution of the Bardeen equation~\cite{Bardeen:1980kt,Mukhanov:2005sc},
\bme{
    \label{eq: Bardeen}
    \Phi_{k}^{\prime\prime}(\eta)+3\mathcal{H}(1+\cs^2)\Phi_{k}^\prime(\eta) \\
    +\bqty{\cs^2k^2+3\mathcal{H}^2(\cs^2-w)}\Phi_{k}(\eta) = 0,
}
with the initial condition $\Phi_k(\eta\to0)\to1$ and $\Phi_k'(\eta\to0)\to0$.
During a pure radiation epoch $\Phi_{k,\rad}(\eta) = 9/(k \eta)^2\left[ \sin(k \eta /\sqrt{3})/(k \eta / \sqrt{3}) - \cos(k \eta / \sqrt{3}) \right]$. To accurately calculate the induced \acp{GW}, we have made a new numerical code using \textit{Julia} language~\cite{bezanson2017julia}, which is publicly available in \cite{codigo_albert}. We give details in the supplemental material.

\begin{figure} 
    \centering
    \includegraphics[width=0.40\textwidth]{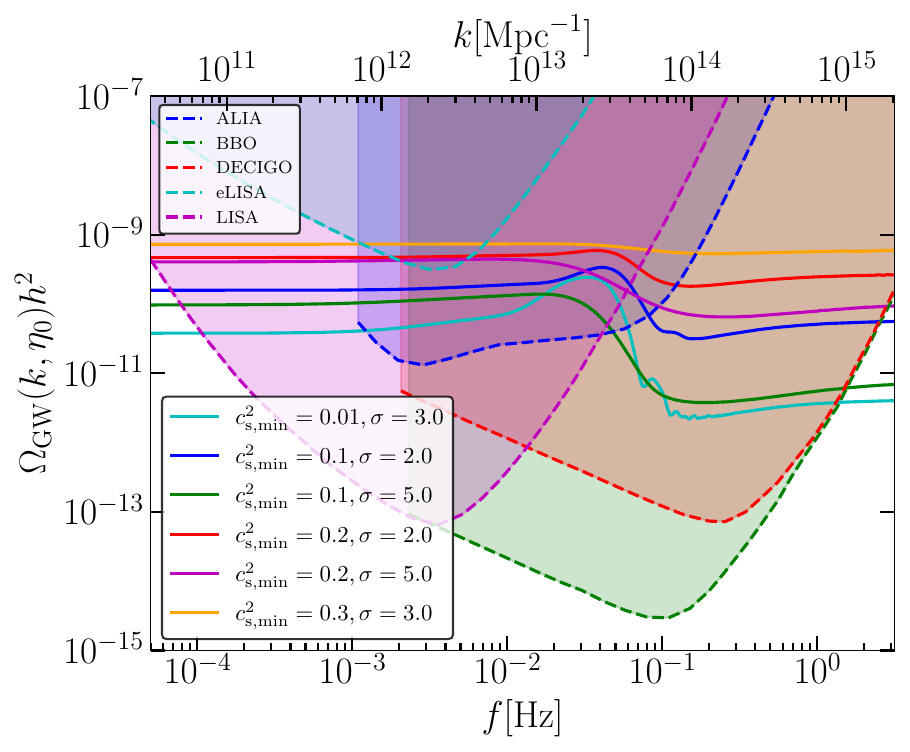}
    \includegraphics[width=0.40\textwidth]{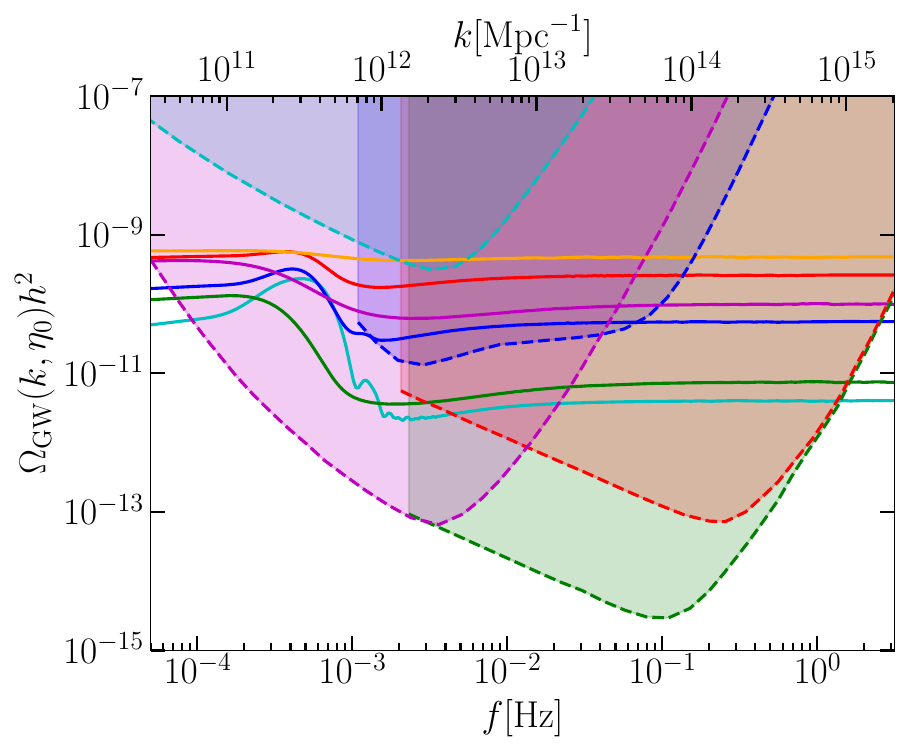}
    \caption{Scalar-induced \acp{GW} from the \ac{SC} for different cases. The top panel corresponds to model \textbf{A} and the bottom to model \textbf{B}. The frequency is related to the wave number as $f=k/2 \pi$. The colored regions indicate sensitivity curves of different planned space-based \acp{GW} interferometers \cite{2015CQGra..32a5014M,Caprini:2019egz,Kawamura:2020pcg} (see the legend).}
    \label{fig:induced_gws}
\end{figure}

The predicted \ac{GW} signature of the \ac{SC} corresponding to the mass function in Fig.~\ref{fig:PBH_mass_function} is shown in Fig.~\ref{fig:induced_gws}. We numerically relate the horizon mass scale (see Fig.~\ref{fig:eq_of_state}) with the wave mode at horizon reentry using the numerical fit shown in Fig.~S4 in the supplemental material. One finds that the crossover effects basically appear as a jump between two levels in the spectrum of \acp{GW}, which is because the smaller $w$ more dilutes the relative energy density of subhorizon \acp{GW} to the background. The differences between the two levels (comparing the values of $\Omega_\GW$ for $k \gg k_{\star}$ and $k \ll k_{\star}$) are given by the factor $a_{\rm sh} \mathcal{H}_{\rm sh}/a_\uf \mathcal{H}_\uf$ in Eq.~\eqref{eq:omega_gw_k_eta0} where $k_{\star}=\mathcal{H}(\eta_{\star})$ with $\eta_{\star}$ the time when $\cs^2\qty(\rho(\eta_{\star}))=\csmin^2$.
In particular, the jump is more than one order of magnitude for the case $\csmin^2=0.1$ and $\sigma=5.0$. On the other hand, when the \ac{SC} has a small softening (basically a pure radiation period), the result of Ref.~\cite{2018PhRvD..97l3532K} is recovered in the limit $w,\cs^2\rightarrow 1/3$. For illustrative purpose of comparison with the related previous work~\cite{Escriva:2022yaf}, 
we also demonstrate the case $\csmin^2=0.01$ and $\sigma=3.0$, which would correspond to the case near 
the critical point between SC and the first-order PT.

The differences between the models \textbf{A} and \textbf{B} basically shift the signal of the induced \acp{GW}, but it has an important implication on the detectability of future space-based planned \acp{GW} interferometers. For the model \textbf{A} (compatible with \acp{PBH} being the dark matter), the \ac{GW} signature associated with the \ac{SC} lies on the frequency range of DECIGO/BBO and partially on LISA. Instead, for the model \textbf{B}, it would lie on the LISA frequency range. 

\section{Conclusions}
%%%%%%%%%%%%%%%%%%%%%%%%%%%%%%%%%%%%%%%%%%%%%%%%%%%
%%%%%%%%%%%%%%%%%%%%%%%%%%%%%%%%%%%%%%%%%%%%%%%%%%%
In this \textit{Letter}, we have numerically estimated the effect of a hypothetical softening of $w$ and $\cs^2$ caused by an \ac{SC} beyond \ac{SM} on the \ac{PBH} mass function and the \acp{GW} spectrum induced by sufficiently large curvature fluctuations. The \ac{PBH} production can be enhanced by an \ac{SC} because the softening of $w$ and $\cs^2$ reduces the formation threshold, which is simulated in numerical relativity (Fig.~\ref{fig:thresholds}).
Consequently, the \ac{PBH} mass function can have a sharp peak as shown in Fig.~\ref{fig:PBH_mass_function} even from the scale-invariant primordial power spectrum~\eqref{eq:ps}. If the \ac{SC} scale $\rho_\uc$ is a few PeV, \ac{PBH} can explain whole \ac{DM} consistently with all the existing observational constraints.

Significant feature in the \ac{GW} spectrum can be evidence of this scenario.
Depending on the mass scale of the crossover, we can detect the ``jump" in the signal in the different \ac{GW} interferometers. In that case, the constant signal before and beyond the jump can be used to infer that the power spectrum of the curvature fluctuation is flat type, and the jump can be associated with the change in time of $w$ and $\cs^2$. In particular, we have shown two representative examples (Fig.~\ref{fig:induced_gws}): i) for $\rho_\uc^{1/4}\approx\SI{1.7e6}{GeV}$ (PBH-DM with the mass $\approx 10^{-13} M_{\odot}$), we find the jump in the signal can be fully detected in the DECIGO/BBO frequencies range and partially in LISA, ii) for $\rho_\uc^{1/4}\approx\SI{17}{TeV}$ corresponding to $\approx 10^{-9} M_{\odot}$, the jump can be detected in the LISA frequency range. Therefore induced GWs can allow us to identify and distinguish that a peaked PBH mass function is caused by a broad power spectrum in the presence of a softening of $w,\cs^2$ in comparison with a peaked power spectrum.

Our results bring a novel motivation for testing new physics at a few TeV--PeV scale giving rise to a significant crossover using induced GWs. Future \ac{GW} interferometers will be essential to test the scenario; see Ref.~\cite{Escriva:2024ivo} for a recent work in this direction for LISA. Future directions could explore non-Gaussianities' impact on both the \ac{PBH} mass function~\cite{Bullock:1996at,Byrnes:2012yx,2021JCAP...10..053K,Escriva:2023uko} and the density shape of \acp{GW}~\cite{Atal:2021jyo,Adshead:2021hnm,Abe:2022xur,Li:2023qua,Li:2023xtl}. Additionally, investigating the $\Omega_\GW$ spectrum with various power spectrum shapes and SC modulations would be intriguing.

%%%%%%%%%%%%%%%%%%%%%%%%%%%%%%%%%%%%%%%%%%%%%%%%%%%
\acknowledgments
A.E. acknowledges support from the JSPS Postdoctoral Fellowships for Research in Japan (Graduate School of Sciences, Nagoya University). 
Y.T. is supported by JSPS KAKENHI Grant
No.~JP21K13918. C.Y. is supported in part by JSPS KAKENHI
Grant Nos. 20H05850 and 20H05853.

%%%%%%%%%%%%%%%%%%%%%%%%%%%%%%%%%%%%%%%%%%%%%%%%%%%
\newpage
\appendix

%%%%%%%%%%%%%%%%%
%%%%%%%%%%%%%%%%%
\onecolumngrid

\section*{Appendix}

\subsection{Numerical computation of the induced \acp{GW}}

We summarize our numerical procedure for the computation of the induced \acp{GW} as follows: i) We find the corresponding wave mode that reenters the horizon at the corresponding $\csmin^2$ scale (namely $k_{\star}$ with $k_{\star}= \mathcal{H}(\eta_{\star})$), and from that scale we consider a range of modes $k \in [10^{-3},10^4] k_{\star}$ in log scale, for what we expect that the \ac{SC} will significantly affect the induced \acp{GW}. ii) For a given mode $k$, we obtain the numerical solution of the homogeneous Green functions $g_{1k}$,$g_{2k}$ (solving Eq.~(13)) using as initial condition the analytical solution when the modes $k$ are well outside the horizon ($\eta_\mathrm{initial} \approx 10^{-2}/k$) during the radiation-dominated epoch, and solved until the modes are well inside the horizon ($\eta_\mathrm{sh} \approx \num{4e3} /k$). iii) For each $k$ considered in the previous step, we solve Eq.~(14) for a range of modes in log-scale $k_{j} \in [10^{-2},10^{2}] \, k$ (the modes $k_{l} \sim k$ will give us already the dominant contribution) with similar $\eta_{\rm initial} \approx 10^{-2}/k_j$ and $\eta_{\rm sh} \approx \num{4e3} /k_j$ as in the previous step, and we store the numerical solution. iv) We make combinations between the modes $k_1$ and $k_2$ following the integral domain of Eq.~(10) and we compute the corresponding kernel functions of Eqs.~(11). We make the integration of Eq.~(10) using the trapezoidal rule approximation with $\sim 500 \times 500$ points of integration in the domain. To compute the time average of the \ac{GW} power spectrum, we follow the practical approach of Refs.~\cite{Abe:2020sqb,Abe:2023yrw}, which consists of taking into account that the integration kernel oscillates only by the mode function $g_{1k}(\eta)$ and $g_{2k}(\eta)$ at the evaluation time $\eta_{\rm sh}$ (since the scalar perturbation is damped enough well after the horizon cross). Then, the time average square of the kernel can be computed as 
\begin{equation}
\label{eq:simplification}
    \overline{I^2(k,k_1,k_2,\eta)}\simeq I_2^2(k,k_1,k_2,\eta)\overline{g_{1k}^2(\eta)} 
    -2I_1(k,k_1,k_2,\eta)I_2(k,k_1,k_2,\eta)\overline{g_{1k}(\eta)g_{2k}(\eta)} 
    +I_1^2(k,k_1,k_2,\eta)\overline{g_{2k}^2(\eta)},
\end{equation}
where $I_l^2(k,k_1,k_2,\eta)$ is given by Eq.~(11) but substituting Green's function by Eq.~(12) and splitting the $g_{lk}$ ($l=1$ or $2$) contributions. The time average of the quantities in Eq.~\eqref{eq:simplification} is defined as,
\begin{equation}
\label{eq:time_average}
    \bar X (\eta)= \frac{1}{T} \int_{\eta-T}^{\eta} X(\tilde{\eta}) \dd{\tilde{\eta}}
\end{equation}
being $T$ the period of the oscillations when the modes $k$ are well inside the horizon. We practically integrate Eq.\eqref{eq:time_average} from the initial time $\eta_{\rm initial}$ until $\eta_{\rm sh}$, which makes the numerical integration easier. This gives equivalent results in the limit when $\eta_{\rm sh} \rightarrow \infty$, which holds in our case with an inappreciable difference in the results with the resolution used.

We check the convergence and accuracy of our numerical result comparing with the analytical value of $\Omega_{\GW} \approx 0.822/\mathcal{A}^2$~\cite{2018PhRvD..97l3532K} for the case of radiation $w=\cs^2=1/3$, for what we find a relative deviation of $\sim \mathcal{O}(0.01\text{--}0.1 \%)$ \footnote{Increasing the number of points for the numerical integration and the final time of evaluation can be observed that the deviation decreases, although the numerical computation is much more expensive. We set $\sim \mathcal{O}(0.1 \%)$ as the limit of accuracy of our numerical results for the $\Omega_{\GW}$, which is already more than enough for our purposes}. This small deviation can be attributed to the highly oscillatory behavior of the kernel function when making the numerical integration.

%%%%%%%%%%%%%%%%

\subsection{Numerical simulations of \ac{PBH} formation}

We briefly review the procedure we have followed to make numerical simulations of the gravitational collapse of super-horizon curvature fluctuations, the threshold values begin shown in Fig.~2 as the final output of the simulations. We follow the approach done in Refs.~\cite{Escriva:2022bwe,Escriva:2022yaf}, which is based on an updated version of the numerical code developed in Ref.~\cite{Escriva:2019nsa}. 

To study the formation of \acp{PBH} from the collapse of a relativistic fluid in spherical symmetry \footnote{Deviations from sphericity could become important in a situation with a very soft constant equation of state~\cite{Kokubu:2018fxy}. In our case, we will consider a transition from radiation with $w$ and $\cs^2$ that becomes softer during a specific narrow period. Therefore, we expect our estimation following spherical symmetry to be realistic since we still consider high peaks.} using the comoving gauge, we solve numerically the \ac{MS} equations~\cite{Misner:1964je}, which are the Einstein equations assuming an energy-momentum tensor given by a perfect fluid as $T^{\mu \nu} = (p+\rho)u^{\mu}u^{\nu}+pg^{\mu\nu}$, with $p=w \rho$. The spacetime metric in spherical symmetry is given by,
\begin{equation}
    \label{2_metricsharp}
    \dd{s^2} = -A(r,t)^2 \dd{t^2}+B(r,t)^2 \dd{r^2} + R(r,t)^2 \left( \dd{\theta^2}+\sin^2(\theta) \dd{\phi^2}  \right).
\end{equation}
where $A$ is the lapse function and $R$ is the areal radius. Then considering a time-dependent equation of state $w(\rho)$, the \ac{MS} equations reads as,
%%%%%%%%%%%%%%%%%%%%%%%%%%%
%\begin{align}
\beae{ 
    \dot{U} &= -A\left[\frac{c^{2}_{\rm s}(\rho)}{1+w(\rho)}\frac{\Gamma^2}{\rho}\frac{\rho'}{R'} + \frac{M}{R^{2}}+4\pi R w(\rho) \rho \right], \\
    \dot{R} &= A U, \\
    \dot{\rho} &= -A \rho \left[1+w(\rho)\right] \left(2\frac{U}{R}+\frac{U'}{R'}\right), \\
    \dot{M} &= -4\pi A w(\rho) \rho U R^{2}, \\ 
    A' &=  -A \frac{\rho'}{\rho} \frac{c^{2}_{\rm s}(\rho)}{1+w(\rho)}, \\
    M' &= 4 \pi \rho R^{2} R',
    \label{eq:MS_equations}
}
%\end{align}
where a dot denotes time derivatives with respect to the cosmic time $t$ and a prime the derivatives with respect to the radius $r$. The sound speed $\cs^2$ is defined %as 
by $\cs^2 = \partial p /\partial \rho $ as in Eq.~(1). The last equation is the Hamiltonian constraint, which is used to check the validity and accuracy of the simulations. $U$ is the radial component of the four-velocity associated with an Eulerian frame (not comoving), which measures the radial velocity of the fluid with respect to the origin of the coordinates. The \ac{MS} mass $M(r,t)$ is defined as $M(R) \equiv \int_{0}^{R} 4\pi \tilde{R}^{2} \rho \dd{\tilde{R}}$ which is related to $\Gamma$, $U$, and $R$ though the constraint $\Gamma = \sqrt{1+U^2-2M/R}$, also given by $\Gamma= R'/B$.

We use gradient expansion approximation approach~\cite{1990PhRvD..42.3936S} to set up the initial condition at the beginning of the simulation, once the fluctuations are at super-horizon scales following Refs.~\cite{2007CQGra..24.1405P,2014JCAP...01..037N,2015PhRvD..91h4057H}. In particular, we consider in the gradient expansion method the ratio between the length-scale of the cosmological horizon $R_H$ and one of the fluctuations $R_\um = a r_\um \ue^{\zeta(r_m)}$, defining a parameter $\epsilon(t) = R_{H}(t)/R_{\um}(t)$ that should be much smaller than one $\epsilon \ll 1$ once the fluctuation is at super-horizon scales. In particular, we take $\epsilon(t_\ui) \lesssim 0.1$, $t_\ui$ being the initial time when we start the simulation. 

\begin{figure} %[t]
    \centering
    \includegraphics[width=2.3 in]{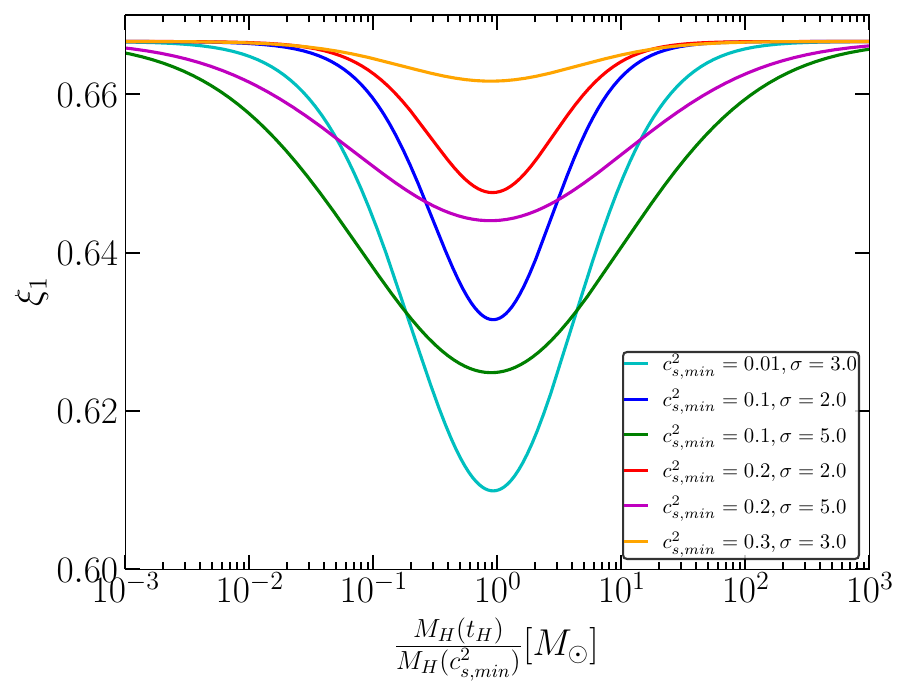}
    \includegraphics[width=2.3 in]{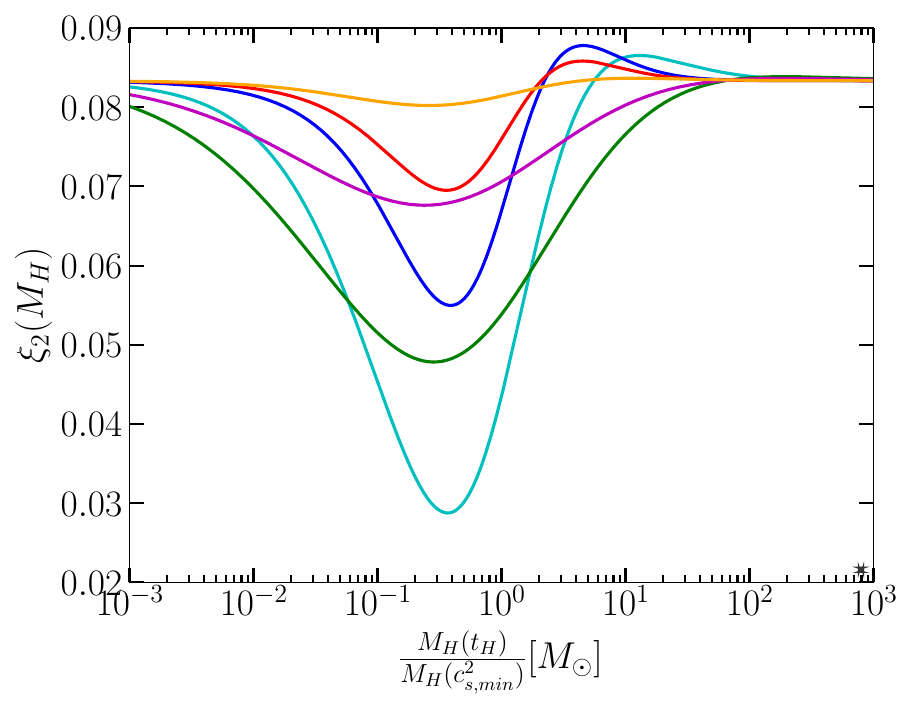}
    \includegraphics[width=2.3 in]{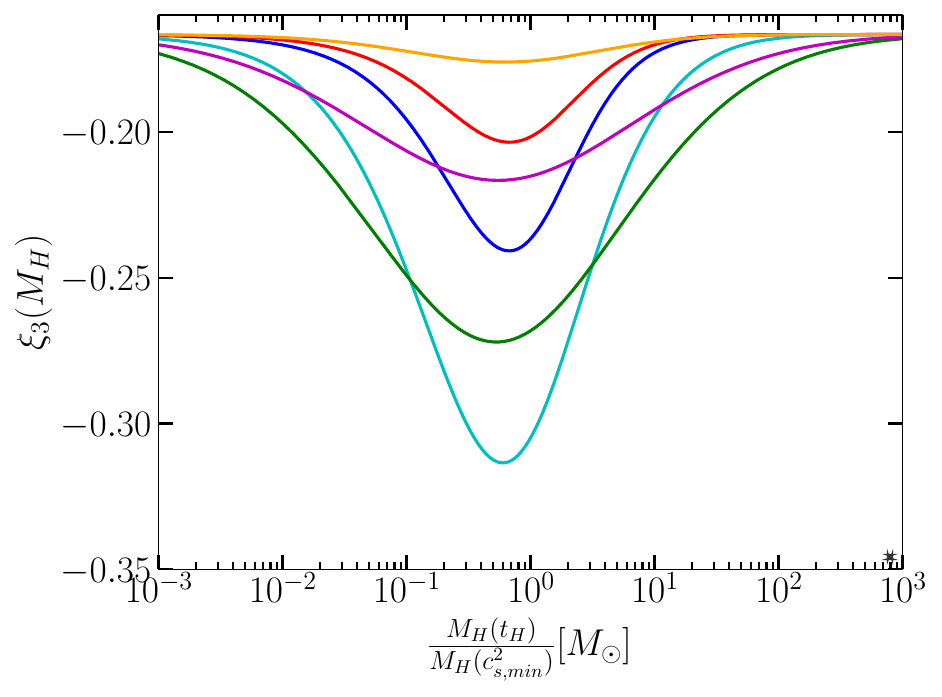}
    \caption{Numerical solutions of $\xi_{1}$, $\xi_{2}$, and $\xi_{3}$ for the different \ac{SC} templates.}
    \label{fig:initial_functions}
\end{figure}

From the spacetime metric of Eq.~(5) and applying the gradient expansion method into Eqs.~\eqref{eq:MS_equations}, the initial conditions for non-constant equations of state $w(\rho)$ are shown in Ref.~\cite{Escriva:2022bwe}. %, and applying 
Applying a convenient change of coordinates from $K(\tilde{r})$ to $\zeta(r)$ given by Ref.~\cite{2015PhRvD..91h4057H}, we then obtain:
%
%\begin{align}
\beae{
    \label{2_expansion}
    A(r,t) &= 1+\epsilon^2(t), \\
    R(r,t) &= a(t)r  \ue^{\zeta(r)}(1+\epsilon^2(t) \tilde{R}), \\ 
    U(r,t) &= H(t) R(r,t) (1+\epsilon^2(t) \tilde{U} ),\\ 
    \rho(r,t) &= \rho_{\ub}(t)(1+\epsilon^2(t)\tilde{\rho}), \\ 
    M(r,t) &= \frac{4\pi}{3}\rho_{\ub}(t) R(r,t)^3 (1+\epsilon^2(t) \tilde{M} ), 
}
%\end{align}
%
%\begin{align}
\beae{
    \phi_{\rho}(r) &= -\frac{1}{3 r} \ue^{-2 \zeta(r)} \left[ 2 r \zeta''(r)+ \zeta'(r) (4+r \zeta'(r))   \right], \\
    \phi_{U}(r) &= -\frac{1}{r} \ue^{-2 \zeta(r)} \zeta'(r) \left[2+ r \zeta'(r) \right] , 
}
%\end{align}
%Then, the hydrodynamic variables at super-horizon scales at the time $t_0$ are,
%\begin{align}
\beae{
    \label{eq:2_perturbations}
    \tilde{\rho}&= \phi_{\rho}\, \xi_{1}(\rho_{\ub,\ui}) \ue^{2 \zeta(r_m)} r^2_\um, \\
    \tilde{U} &= \frac{1}{2}\phi_{U} (\xi_{1}(\rho_{\ub,\ui})-1)\ue^{2\zeta(r_m)}r^2_{\um}, \\
    \tilde{A} &= -\frac{c^2_s(\rho_{\ub,\ui})}{1+w(\rho_{\ub,\ui})}\tilde{\rho}, \\
    \tilde{M} &= 2\frac{\xi_{1}(\rho_{\ub,\ui})}{\xi_{1}(\rho_{\ub,\ui})-1} \tilde{U}, \\
    \tilde{R} &= -\frac{\xi_{2}(\rho_{\ub,\ui})}{\xi_{1}(\rho_{\ub,\ui})}\tilde{\rho}+\frac{\xi_{3}(\rho_{\ub,\ui})}{\xi_{1}(\rho_{\ub,\ui})-1}\tilde{U},
}
%\end{align}
where $\rho_{\ub,\ui}= \rho_\ub(t_\ui)$ and $\xi_{1}(\rho_{\ub})$, $\xi_{2}(\rho_{\ub})$, and $\xi_{3}(\rho_{\ub})$ are functions of the energy density of the \ac{FLRW} background, obeying the following differential equations:
%\begin{align}
\beae{
    \label{eq:initial_functions}
    \dv{\xi_{1}(\rho_{\ub})}{\rho_{\ub}} &= -\frac{1}{2 \rho_{\ub}} + \frac{5+3w(\rho_{\ub})}{2\left[1+w(\rho_{\ub})\right]} \frac{\xi_{1}(\rho_{\ub})}{3 \rho_{\ub}}, \\
    \dv{\xi_{2}(\rho_{\ub})}{\rho_{\ub}} &= -\frac{\cs^{2}(\rho_{\ub})}{3\left[1+w(\rho_{\ub})\right]^{2}}  \frac{\xi_{1}(\rho_{\ub})}{\rho_{\ub}} + \frac{\left[1+3w(\rho_{\ub})\right]}{3\left[1+w(\rho_{\ub})\right]}\frac{\xi_{2}(\rho_\ub)}{\rho_\ub}, \\
    \dv{\xi_{3}(\rho_{\ub})}{\rho_{\ub}} &= \frac{-1}{3\left[1+w(\rho_{\ub})\right]} \frac{\left[\xi_{1}(\rho_{\ub})-1 \right]}{\rho_{\ub}} + \frac{\left[1+3w(\rho_{\ub})\right]}{3\left[1+w(\rho_{\ub})\right]}\frac{\xi_{3}(\rho_{\ub})}{\rho_{\ub}}.
}
%\end{align}
%
Eqs.~\eqref{eq:initial_functions} are solved numerically using the analytical templates of Eqs.~(1) and (2) for $\cs^{2}(\rho_{\ub})$ and $w(\rho_{\ub})$, and with the initial conditions such that $\dv{\xi_1(\rho_{\ub,\rad})}{\rho_{\ub}} = \dv{\xi_{2}(\rho_{\ub,\rad})}{\rho_{\ub}} = \dv{\xi_{3}(\rho_{\ub,\rad})}{\rho_{\ub}}=0$ where $\rho_{\ub,\rad}= \rho_{\ub}(t_{\rad})$ and $t_{\rad}$ is the time where we have $w=\cs^2=1/3$. The numerical solution is shown in Fig.~\ref{fig:initial_functions}.

The formation of a \ac{PBH} for a given initial condition can be inferred from the dynamics of perturbations
that continue growing (i.e, which do not dissipate) after entering the horizon until the formation of an apparent horizon~\cite{1965PhRvL..14...57P}. In spherical symmetry, this condition is satisfied when $2M=R$. Following a numerical bisection, the critical value $\mu_{2,\uc}$ can be obtained. We make the simulations to obtain the thresholds with a resolution $\mathcal{O}(10^{-2}\%)$ as in Ref.~\cite{Escriva:2022bwe}, which means an absolute resolution of $\mathcal{O}(10^{-4})$ depending on the \ac{SC} template considered.

\begin{figure} %[h]
    \centering
    \includegraphics[width=2.3 in]{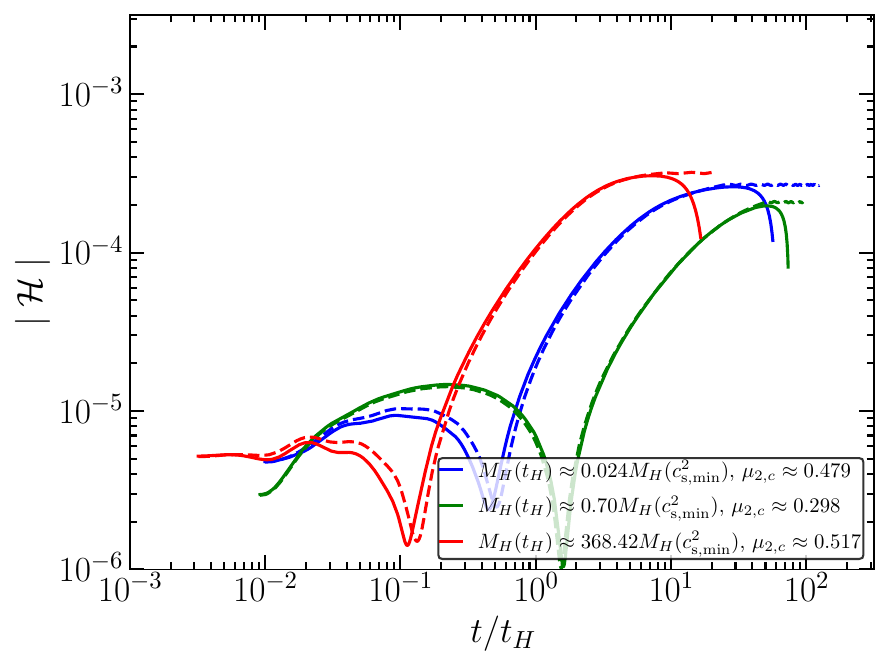}
    \includegraphics[width=2.3 in]{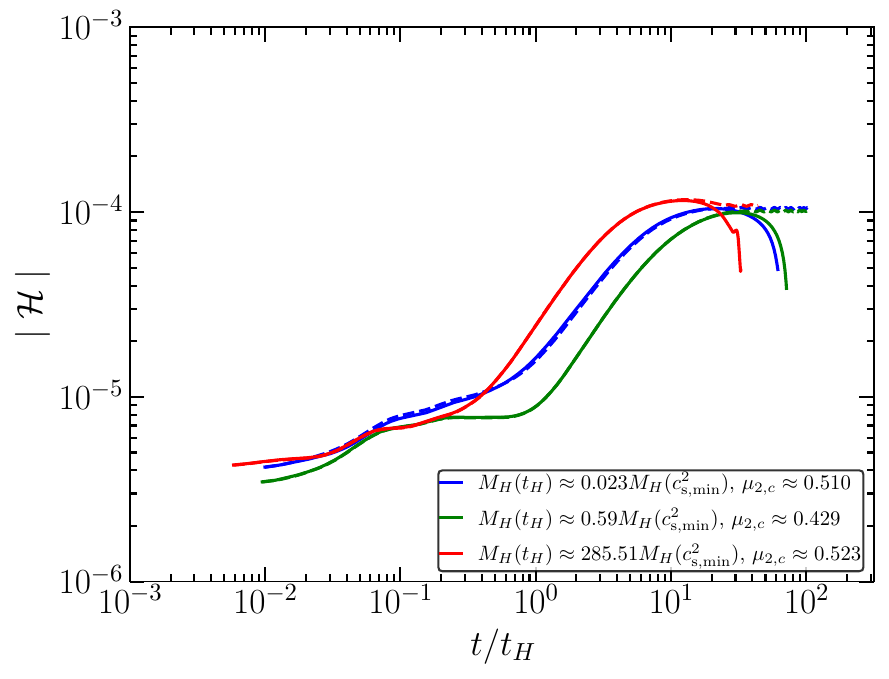}
    \includegraphics[width=2.3 in]{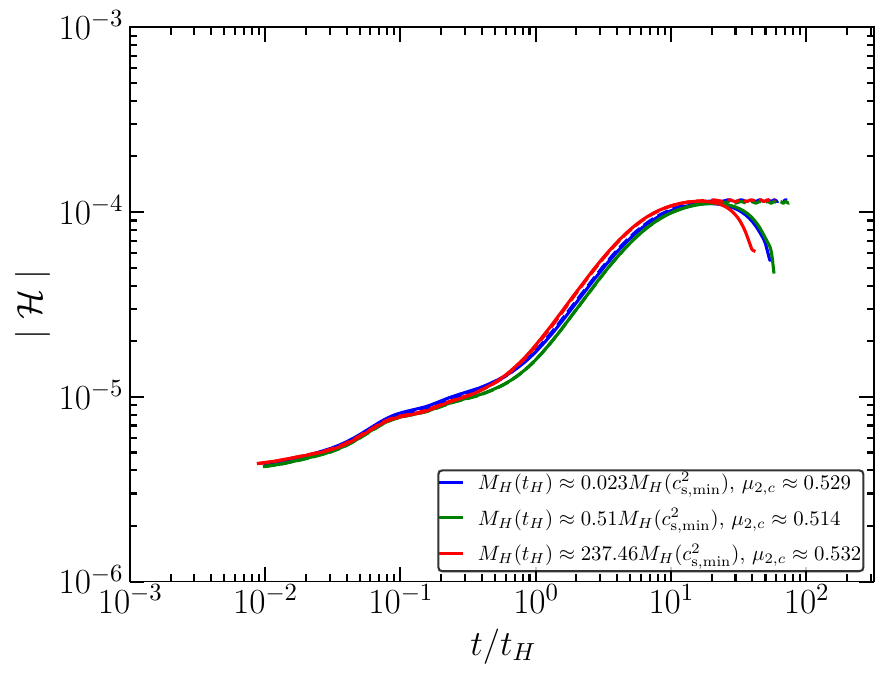}
    \includegraphics[width=2.3 in]{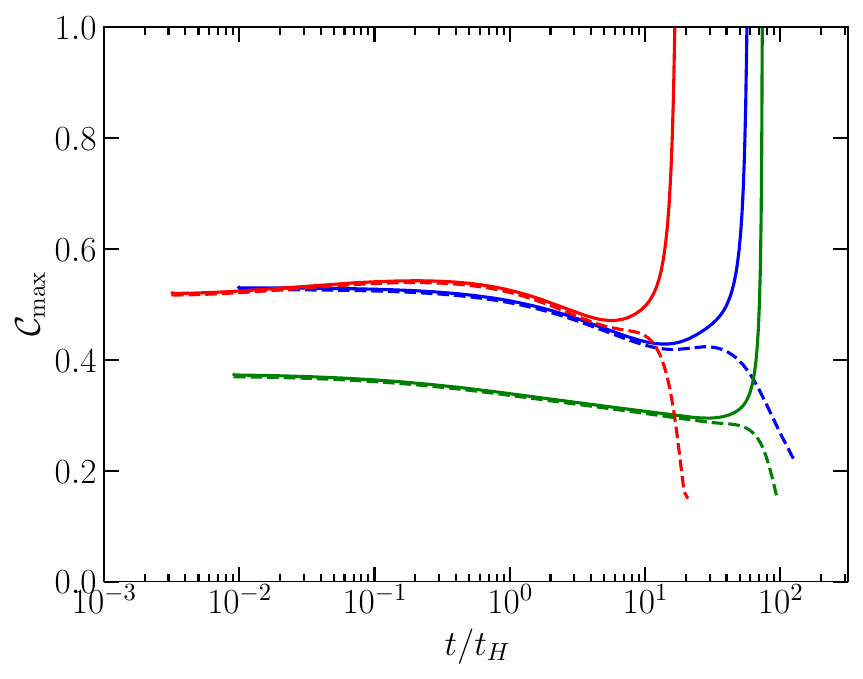}
    \includegraphics[width=2.3 in]{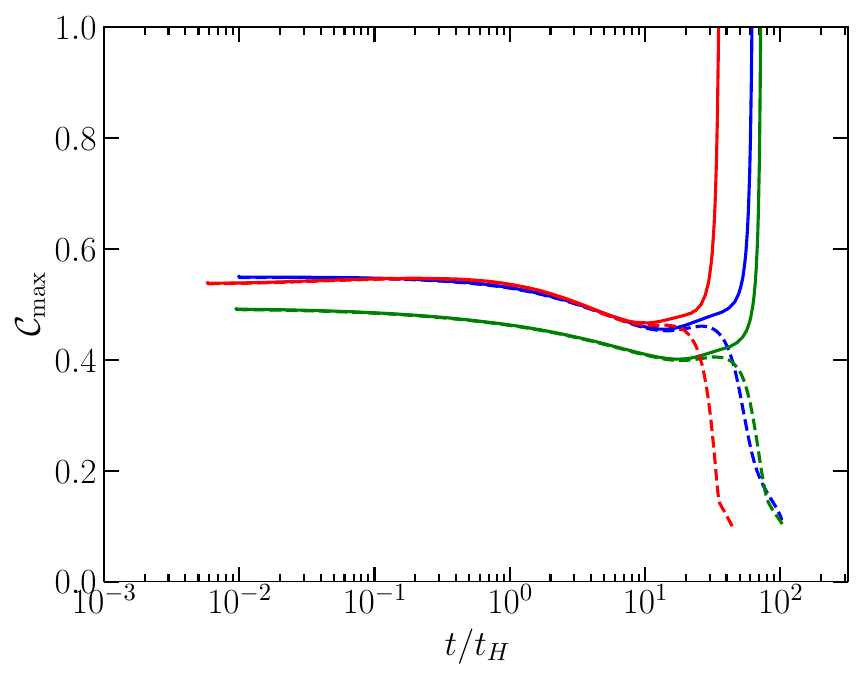}
    \includegraphics[width=2.3 in]{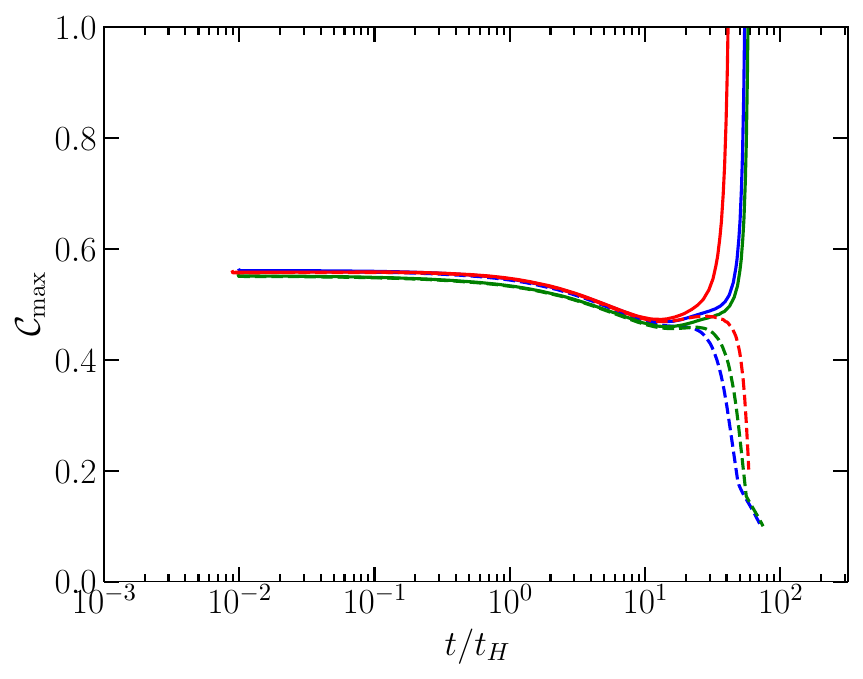}
    \caption{Hamiltonian constraint evolution (top panels) and the time evolution of the maximum of the compaction function (bottom panels) for different \ac{SC} templates and $M_{H}(t_H)$. The solid line corresponded to cases with apparent horizon formation and dashed lines cases where the cosmological fluctuation is dispersed on the \ac{FLRW} background (no black hole formation). Left panels: $\csmin^2 = 0.1$ and $\sigma=5$, middle panels: $\csmin^2 = 0.2$ and $\sigma=5$, and right panels: $\csmin^2 = 0.3$ and $\sigma=5$. The cases shown corresponds to $\mu_{2} = \mu_{2,\uc} \pm \num{5e-4}$.}
    \label{fig:H_constraint_apendix}
\end{figure}

%%%%%%%%%%%%%%%%%%%%%%%%%
%%%%%%%%%%% accuracy simulations %%%%%%%%%
To check the accuracy of our simulations, we use the Hamiltonian constraint equations $M' \equiv 4 \pi R' R^2 \rho$ to define the following quantity,

%we use the Hamiltonian constrain equation $M' \equiv 4 \pi R' R^2 \rho$ to define the quantity,
\begin{equation}
    \mathcal{H} \equiv \frac{M'_\mathrm{num}-M'_\mathrm{def}}{M'_\mathrm{def}} = \frac{M'_\mathrm{num}/R'_\mathrm{num}}{4 \pi \rho_\mathrm{num}R^2_\mathrm{num}}-1,
\end{equation}
where the square norm is given by,
\begin{equation}
    \abs{\mathcal{H}} \equiv \frac{1}{N_\mathrm{cheb}} \sqrt{\sum_{i=1}^{N} \left(\frac{M'_i/R'_i}{ 4\pi \rho_i R^2_i}-1 \right)^2}.
    \label{eq:constraint_numerical}
\end{equation}
The sub-index $i$ refers to each grid point, and $N_\mathrm{cheb}$ is the total number of grid points. To ensure the accuracy of the simulations, $\abs{\mathcal{H}}$ should be much smaller than one. Examples of the convergence of our simulations are shown in Fig.~\ref{fig:H_constraint_apendix} for the calculation of the threshold values $\mu_{2,\uc}$ for different \ac{SC} templates. We use the maximum of the compaction function $\mathcal{C}_{\umax}$ to infer black hole formation (when $\mathcal{C}_{\umax} \approx 1$) or dispersion of the fluctuation on the \ac{FLRW} background (when $\mathcal{C}_{\umax}$ continuously decreases in time)~\cite{Escriva:2019nsa}.

% Hamiltonian_constraintcs_min_0.1sigma_5.0.pdf

\subsection{\ac{PBH} abundance estimation with peak theory}
\label{sec:abundance_pbh_appendix}

In this section, we give details about the statistical estimation of \ac{PBH} abundances we have followed and the approximations considered. We refer the reader to Refs.~\cite{1986ApJ...304...15B,pic_original,Yoo:2020dkz,2021JCAP...10..053K} for much more details.

We make statics on counting peaks of the Laplacian of the curvature fluctuation $\Delta \zeta$. This has the advantage of decoupling the large-scale environmental effect in the estimation of \ac{PBH} abundance, which allows the correct estimate of the abundance in the case of a flat power spectrum through the introduction of a window function, in comparison with the previous approach~\cite{pic_original}.

Let us consider that the curvature fluctuation $\zeta$ follows a Gaussian distribution, with a power spectrum defined as
\begin{equation}
    \expval{\zeta(\bold{k}) \zeta^{*}(\bold{k'})} = \frac{2 \pi^2}{k^3} \mathcal{P}_{\zeta}(k) (2 \pi)^2 \delta(\bold{k}-\bold{k'})
\end{equation}
where $\bold{k}$ and $k$ are the wave-mode vector and its modulus respectively. Then according to peak theory~\cite{1986ApJ...304...15B} and making an integration over the \textit{typical profile} of $\Delta \zeta$~\cite{Yoo:2020dkz}, the \textit{typical profile} of $\zeta$ is given by $\zeta(r) = \mu_{2} g(r; k_{\bullet})$ with
\begin{equation}
\label{eq:typical_profile_apendix}
g(r; k_{\bullet}) =   \frac{1}{1-\gamma_{3}^2} \left(\psi_{1}+\frac{1}{3}R_{3}^2 \Delta \psi_1\right) -\frac{k_{\bullet}^2}{\gamma_3 (1-\gamma_3^2)}\frac{\sigma_2}{\sigma_4}\left( \gamma_3^2 \psi_1 + \frac{1}{3} R_3^2 \Delta \psi_1  \right)  +  \zeta_{\infty}
\end{equation}
where the last term is an integration constant and $\mu_2$ and $k_\bullet$ are defined by $\mu_2 = -(\sigma^2_1 / \sigma^2_2)\left.\Delta  \zeta \right|_{r=0}$ and $k_{\bullet} = - \left.\Delta \Delta \zeta \right|_{r=0}/ (\left.\Delta  \zeta \right|_{r=0})$. The term $\zeta_{\infty}$ can be considered as a Gaussian distributed variable, with a mean value equal to zero. It can be discarded by making a renormalization as discussed in Ref.~\cite{Yoo:2020dkz}. The statistical parameters in Eq.~\eqref{eq:typical_profile_apendix} used to build $g$ are,
\beae{
\label{eq: stat params}
    &\sigma_n^2=\int\frac{\dd{k}}{k}k^{2n}\calP_g(k), &
    &\psi_n(r)=\frac{1}{\sigma_n^2}\int\frac{\dd{k}}{k}k^{2n}\frac{\sin(kr)}{kr}\calP_g(k), \\
    &\gamma_n=\frac{\sigma_n^2}{\sigma_{n-1}\sigma_{n+1}}, &
    &R_n=\frac{\sqrt{3}\sigma_n}{\sigma_{n+1}}  \, \,\,  \text{for odd $n$.}
    }

Notice that these quantities are not affected by the existence of an SC since they rely on the curvature fluctuation $\zeta$, which is frozen at super-horizon scales. The number density of such a peak in a comoving volume (i.e., the comoving number density of positive extremal points of $g$ in a high peak limit $\nu = \mu_{2}/\sigma_{2} \gg 1$) is furthermore expected statistically as
\begin{equation}
\label{eq: npk mu0 k1}
    n_\pk^{(\mu_2,k_{\bullet})}\dd{\mu_2}\dd{k_{\bullet}}=\frac{2}{3^{3/2} \cdot (2\pi)^{3/2}}\mu_2 k_{\bullet}\frac{\sigma_2^3 \sigma^2_4}{\sigma^4_1 \sigma_3^3}  f\left(\frac{\sigma^2_2}{\sigma^2_1 \sigma_4} \mu_2 k^2_{\bullet} \right)P_1^{(3)}\left(\frac{\sigma_2}{\sigma^2_1}\mu_2,\frac{\sigma^2_2}{\sigma^2_1 \sigma_4  }\mu_2 k^2_{\bullet} \right)\dd{\mu_2}\dd{k_\bullet},
\end{equation}
where
\begin{equation}
    f(\xi)=\frac{1}{2}\xi(\xi^2-3)\left(\erf\left[\frac{1}{2}\sqrt{\frac{5}{2}}\xi\right]+\erf \left[ \sqrt{\frac{5}{2}}\xi\right]\right)  +\sqrt{\frac{2}{5\pi}}\left[\left(\frac{8}{5}+\frac{31}{4}\xi^2\right)\exp\left(-\frac{5}{8}\xi^2\right)+\left(-\frac{8}{5}+\frac{1}{2}\xi^2\right)\exp\left(-\frac{5}{2}\xi^2\right)\right], 
    \label{eq: f}
\end{equation}
and
\begin{equation}
    P_1^{(n)}(\nu,\xi)=\frac{1}{2\pi\sqrt{1-\gamma_n^2}}\exp\left[-\frac{1}{2}\left(\nu^2+\frac{(\xi-\gamma\nu)^2}{1-\gamma_n^2}\right)\right].
    \label{eq:P1}
\end{equation}
Here $\erf (z)$ denotes the error function $\erf(z)=\frac{2}{\sqrt{\pi}}\int^z_0 \ue^{-t^2}\dd{t}$. The amplitude $\mu_2$ is related to the compaction function $\mathcal{C}$~\cite{Shibata:1999zs} at super-horizon scales as
\begin{equation}
    \mu_{2} = \frac{-1+\sqrt{1-3 \mathcal{C}(r_\um)/2}}{r_\um \eval{\partial_{r} g_\um(r;k_{\bullet})}_{r=r_\um}}.
\end{equation}
The number density of \acp{PBH} can be written as  
\begin{equation}
    n_{\PBH} \dd{\ln M} = \left[  \int_{\mu_{2,\uc}}^{\infty}  \dd{\mu_{2}} n_\pk^{(\mu_2,M)} \right] \dd{\ln M},
\end{equation}
where we change the variable $k_{\bullet}$ to $M$ in Eq.~\eqref{eq: npk mu0 k1} using the Jacobian of the transformation $\dd{k_{\bullet}} = \abs{\dv*{\ln M}{k_{\bullet}}}^{-1} \dd{\ln M}$. At this point, we consider three approximations that we use in the next steps: i) The \ac{PBH} mass is given by the mass of the horizon at horizon reentry $M_{H}(t_H)=H^{-1}(t_H)/2$ ($t_H$ being the time of horizon crossing of the comoving scale $r_\um$), which would over-estimate the \ac{PBH} mass slightly since the \acp{PBH} that most contribute to the \ac{PBH} abundance would be $\alpha M_{H}(t_H)$ with a factor $\alpha \in [0.1\text{--}1]$, which depends on the equation of state and the curvature profile \cite{2021JCAP...05..066E}. ii) We neglect the effect of the critical regime on the \ac{PBH} mass. As shown in Ref.~\cite{Yoo:2022mzl} for a pure radiation case, that effect is anyway very small on the total abundance $f_{\PBH}^{\tot}=\int f_{\PBH}(M) \dd{\ln M}$. Moreover, the inclusion of the critical regime would make only a minor change in the shape of the mass function (see, for instance, Ref.~\cite{Escriva:2022bwe}). The dominant contribution comes from the reduction of $\mu_{2,\uc}\qty(M_H(t_H))$. iii) To simplify the computations, we also assume that $M_\ut$ is computed assuming a pure radiation-dominated Universe. This would underestimate the \ac{PBH} mass by a factor $\sim 3$ at most (see Fig.~\ref{fig:horizon_mas}). We expect that the approximations i) and iii) reduce their impact when taken together since they over-estimate/underestimate the \ac{PBH} mass.

Then the abundance of \acp{PBH} $\beta$ at the time of matter-radiation equality (denoted by the sub-index ``0") is given by

\begin{figure} %[h]
    \centering
    \includegraphics[width=2.6 in]{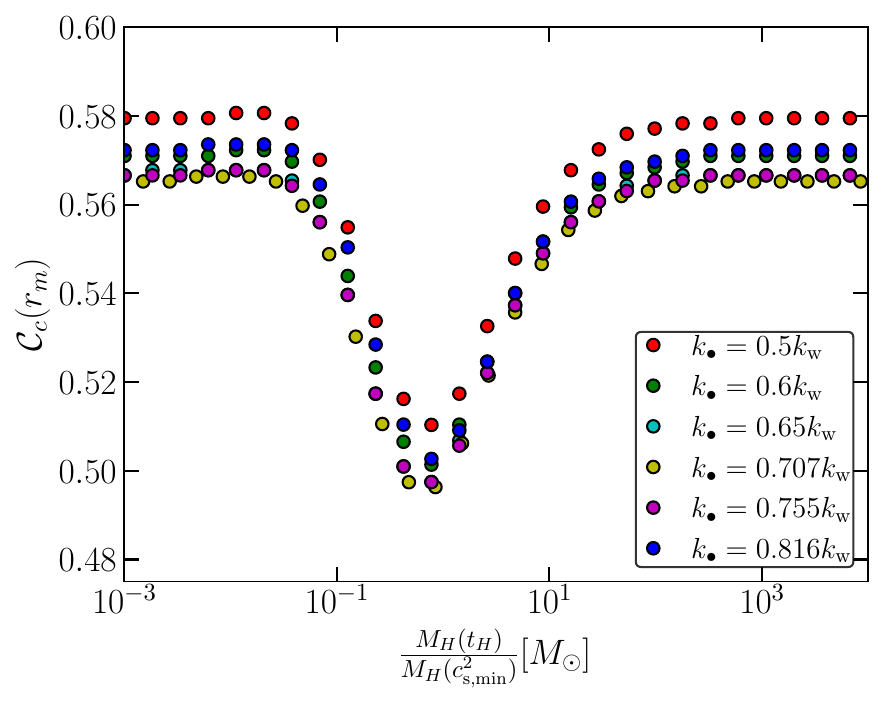}
    \includegraphics[width=2.6 in]{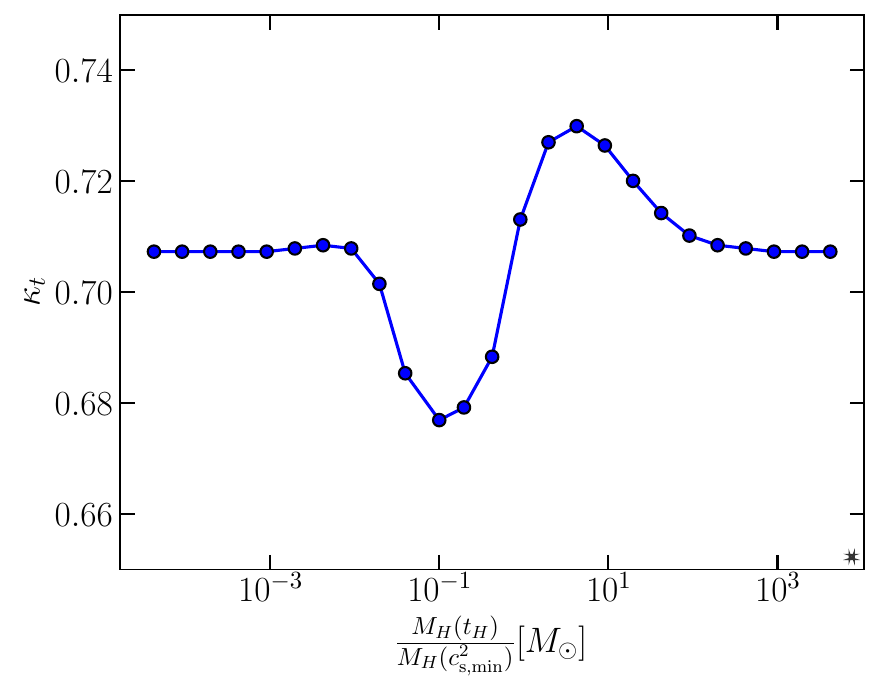}
    \includegraphics[width=2.6 in]{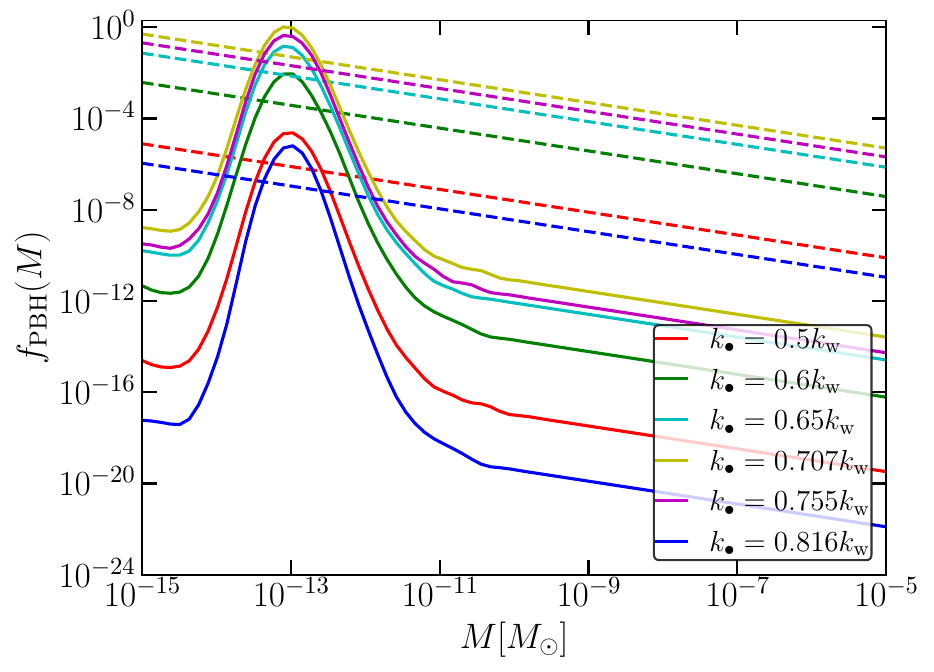}
    \caption{Left-top panel: Threshold values $\mu_{2,\uc}(M_H(t_H))$ for different realisations of $k_{\bullet}$. Right-top panel: Values of $\kappa_t$ that maximises the PBH abundance. Bottom panel: \ac{PBH} mass function using the values of $\mu_{2,\uc}(M_H(t_H))$ from the top-left panel and using the same amplitude $\mathcal{A}$. In all cases, we have considered the \ac{SC} template with $\csmin^2 = 0.2$ and $\sigma= 3.0$.}
    \label{fig:variation_kt}
\end{figure}

\bae{
    \beta_0\dd{\ln M}&=\frac{M n_{\PBH}}{\rho a^3}\dd{\ln M}
    =\frac{4\pi}{3} n_{\PBH}k_{\eq}^{-3}\left(\frac{M}{M_{\eq}}\right)^{3/2} \dd{\ln M} \nonumber \\
    &\bmbe{=\frac{2\times3^{-5/2} k_{\eq}^{-3}}{(2\pi)^{1/2}}
    \frac{\sigma_4^2 \sigma^3_2}{\sigma^4_1 \sigma_3^3}
    \left(\frac{M}{M_{\eq}}\right)^{3/2}
    \Biggl[
    \int^\infty_{\mu_{\rm 2,\uc}}\dd{\mu}_2
    \mu_2 k_\bullet f\left(\frac{\sigma^2_2}{\sigma^2_1 \sigma_4} \mu_2 k_\bullet^2\right) \\
    \times P_1\left(\frac{\sigma_2}{\sigma^2_1}\mu_2,\mu_2 k_\bullet^2 \frac{\sigma^2_2}{ \sigma^2_1 \sigma_4 }\right)
    \abs{\dv{k_\bullet}\ln  r_{\um}+\mu_2 \dv{k_\bullet}g_{\um}}^{-1}
    \Biggr]\dd{\ln M},
    }
    \label{eq:beta_general}
}
%\end{eqnarray}
%
where the scale factor $a$ is written as a function of $M$ as $a=2M^{1/2}M_{\eq}^{1/2}k_{\eq}$. 
%\st{and we neglect the critical behavior of the \ac{PBH} mass assuming that $M = H^{-1}(t_H)/2$, which corresponds to the mass of the cosmological horizon at the time of horizon reentry $t_H$}. 
The integral of Eq.~\eqref{eq:beta_general} with respect to $\mu_{2}$ can be approximated as follows:
%
%\begin{eqnarray}

\begin{multline} 
    \beta_0\dd{\ln M}\simeq
    \frac{2\times3^{-5/2} k_{\eq}^{-3}}{(2\pi)^{1/2}}
    \frac{\sigma_4^2}{\sigma_2\sigma_3^3}
    \left(\frac{M}{M_{\eq}}\right)^{3/2}
    \Biggl[
    \tilde\sigma^2(k_\bullet)
    k_\bullet f\left(\frac{\sigma^2_2}{\sigma^2_1\sigma_4} \mu_2 k_\bullet^2\right) \\
    \times P_1\left(\frac{\sigma_2}{\sigma^2_1}\mu_2,\mu_2 k_\bullet^2 \frac{\sigma^2_2}{\sigma^2_1 \sigma_4 }\right)
    \abs{\dv{k_\bullet}\ln r_{\um}+\mu_2 \dv{k_\bullet}g_{\um}}^{-1}
    \Biggr]_{\mu_2=\mu_{2,\uc}}\dd{\ln M},
    \label{eq:beta_approx}
\end{multline}
%\end{eqnarray}
%
where $\tilde{\sigma}$ is defined as,
\begin{equation} 
    \frac{1}{\tilde{\sigma}^2(k_{\bullet})} = \frac{1}{\sigma^2_2} + \frac{1}{\sigma^2_4 (1-\gamma^2_3)}\left(  k^2_{\bullet}  -\frac{\sigma^2_3}{\sigma^2_2} \right)^2.
\end{equation}
Since $P_1$ given in Eq.~\eqref{eq:P1} has the exponential dependence, we may expect that the value of $\beta_0$ is sensitive to the exponent $-\tilde{\mu}_2^2/2\tilde{\sigma}^2$ (with $\tilde{\mu}_2 = \mu_2 \sigma^2_2/\sigma^2_1$). We can roughly estimate the maximum value of $\beta_0$ at the top of the mass spectrum by considering the value $k_{\ut}$ of $k_{\bullet}$ which minimizes the value of $\tilde{\mu}_{2,c}^{(k_\bullet)}(k_\bullet)/\tilde{\sigma}$, namely
\begin{equation} 
    k_{\ut}:=\argmin_{k_\bullet}\left[\tilde{\mu}_{2,c}^{(k_\bullet)}(k_\bullet)/\tilde{\sigma}(k_\bullet)\right].  
%\footnote{
%\baselineskip5mm
%${\rm argmin}_x f(x)=\{x~|~\forall y (f(y) 
%\geq f(x))\}$. 
%}
\end{equation}
In general, a numerical procedure is needed to determine the value of $k_{\ut}$, which is independent of the amplitude of the power spectrum and only depends on its shape. In our case, we find that $k_{\ut} \approx 0.707 k_\uW$ maximizes the corresponding mass function and with the lowest thresholds $\mathcal{C}_{\uc}(r_\um)$ as shown in Fig.~\ref{fig:variation_kt} for the specific case of $\csmin^2=0.2$ and $\sigma=3$.

In our numerical exploration, we find a variation $\delta k_\ut \approx \pm 0.03$ to the exact value of $k_\ut$ that maximizes the abundance (see top-right panel of Fig.~\ref{fig:variation_kt}). We assume that $k_{\ut} \approx 0.707 k_\uW$ (the value for radiation-dominated Universe) also holds for the other \ac{SC} templates with similar variations of $\delta k_\ut$. Even assuming a possible variation of $k_\ut$ for other \ac{SC}, we expect that the maximum value of the mass function will be determined by $k_{\ut} \approx 0.707 k_\uW$ at the mass scale corresponding to the minimum of $\mu_{2,\uc}$ (as already proven for the case tested in Fig.~\ref{fig:variation_kt}), since it will act as an attractor because for constant $w,c^2_s$ holds that $k_{\ut} \approx 0.707 k_\uW$ maximizes the abundance.

%%%%%%%%%%%%%%%%%
%%%%%%%%%%%%%%%%%%
%%%%%%%%%%%%%%%%%%
%%%%%%%%%%%%%%%%%%

Notice that our choice of $k_\ut$ differs from the \textit{mean profile}, which corresponds to $k_{\bullet} = \sigma_3 / \sigma_2 \approx 0.816 k_\uW$ ($\kappa \approx 0.816$) with $\lambda \approx 3.5$, corresponding to $\zeta = \mu_2 \psi_1$. According to our results, the assumption of the \textit{typical profile} to be coincident with the \textit{mean profile} in the case of a flat power spectrum would substantially underestimate the \ac{PBH} production by several orders of magnitude, fixing the same amplitude of the power spectrum $\mathcal{A}$. In other words, the most statistically relevant curvature profile $\zeta$ is not always the \textit{mean profile} \footnote{It is only exactly true for the monochromatic power spectrum since only a single wave mode is involved in the gravitational collapse. It is also expected to be a very good approximation for a sharply peaked power spectrum.}.

%%%%%%%%%%%%%%%%%
%%%%%%%%%%%%%%%%%%
%%%%%%%%%%%%%%%%%%
%%%%%%%%%%%%%%%%%%

When computing the \ac{PBH} abundance, several realizations of the curvature profiles need to be taken into account for the different modulations $k_{\bullet}$. In our case, due to the %computational time-consuming 
time-consuming computation
of finding the numerical threshold $\mu_{2,\uc}(M_H)$ for different $k_{\bullet}$ and templates of the \ac{SC}, we have searched for the most likely statistical realization fixing the value of $k_\ut$, which we expect give us the maximum contribution to the \ac{PBH} abundance according to our approximations. Notice that in Eq.~\eqref{eq:typical_profile_apendix} we could have also included the dispersion in the shapes following Ref.~\cite{1986ApJ...304...15B} as done in Ref.~\cite{2020JCAP...05..022A}, although we expect it to be small for the high peak limit, as shown in Ref.~\cite{2020JCAP...05..022A} for the case of a peaked power spectrum. In addition, the choice of the window function can also affect the results, although we have chosen the top-hat following Ref.~\cite{Yoo:2020dkz}, which gives us the maximum abundance with the minimal set-up. We leave the consideration of all these effects for future research.

Substituting $k_{\ut}$ into $k_\bullet$ in Eq.~\eqref{eq:beta_approx}, we obtain the following rough estimate for the maximum value of $\beta_{0,\umax}$: 
%\begin{eqnarray}
\begin{multline} 
    \beta_{0,\umax}\simeq
    \beta_{0,\umax}^{\uapprox}
    :=
    \frac{2\times3^{-5/2} k_{\eq}^{-3}}{(2\pi)^{1/2}}
    \frac{\sigma_4^2 }{\sigma_2 \sigma^3_3}
    \left(\frac{M_{\ut}}{M_{\eq}}\right)^{3/2}
    \Biggl[
    \tilde{\sigma}^2(k_\bullet)
    k_\bullet f\left(\frac{\sigma^2_2}{\sigma^2_1\sigma_4} \mu_2 k_\bullet^2\right) \\
    \times P_1\left(\frac{\sigma_2}{\sigma^2_1}\mu_2,\mu_2 k_\bullet^2 \frac{\sigma^2_2}{\sigma^2_1 \sigma_4 }\right)
    \abs{\dv{k_\bullet}\ln r_{\um}+\mu_2 \dv{k_\bullet}g_{\um}}^{-1}
    \Biggr]_{k_\bullet=k_{\ut},~\mu_2=\mu_{2,c}^{(k_\bullet)}(k_{\ut})}.
    \label{eq:beta_max}
\end{multline}
%\end{eqnarray}
Finally, the mass function shown \footnote{Notice that the constraints shown in Fig.~(3) apply to the monochromatic power spectrum, which is not necessarily our case having a flat power spectrum. Nevertheless, we take the constraints as an indication.} in Fig.~(3) is computed as $f_\PBH(M\mid k_\uW)=\sqrt{M/M_{\eq}}\beta_{0,\umax}^{\uapprox}$.

As a final remark, notice that in our approach, the use of a transfer function is unnecessary since numerical simulations (starting initially with the fluctuations at super-horizon scales) are already telling us the final output (threshold), and we account for peaks of $\Delta \zeta$ at super-horizon scale when $\zeta$ is frozen and Gaussian randomly distributed, for what we can correctly account for the statistics~\cite{1986ApJ...304...15B}.

Our numerical and theoretical approach using peak theory can be used to study the case of the QCD crossover in the range of $T \approx 200 \textrm{MeV}$ and compare it with the current literature on the topic. We leave that for future research.

\subsection{Mass of the cosmological horizon with the \ac{SC}}

\begin{figure} %[h]
    \centering
    \includegraphics[width=2.7 in]{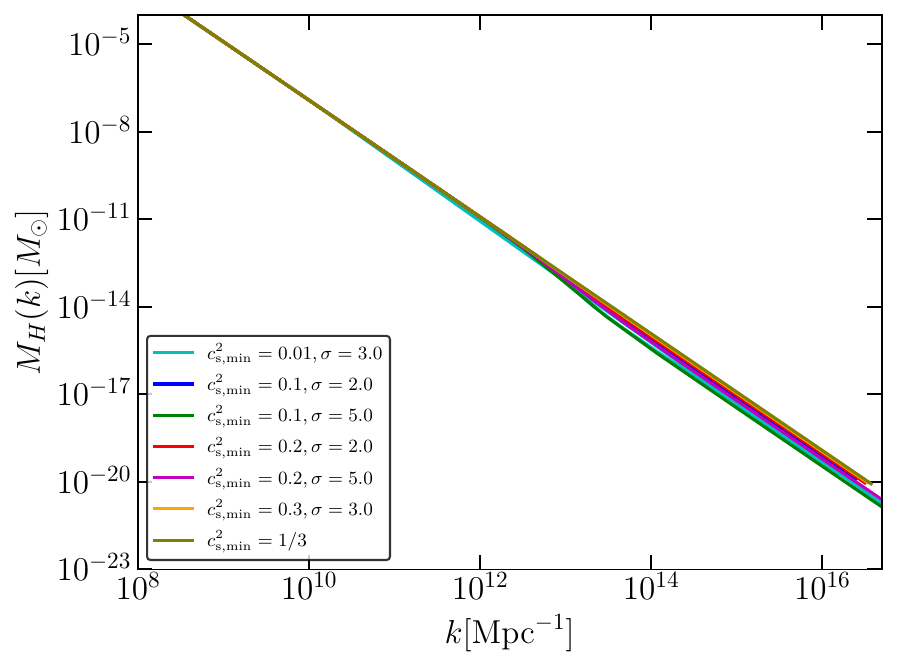}
    \includegraphics[width=2.7 in]{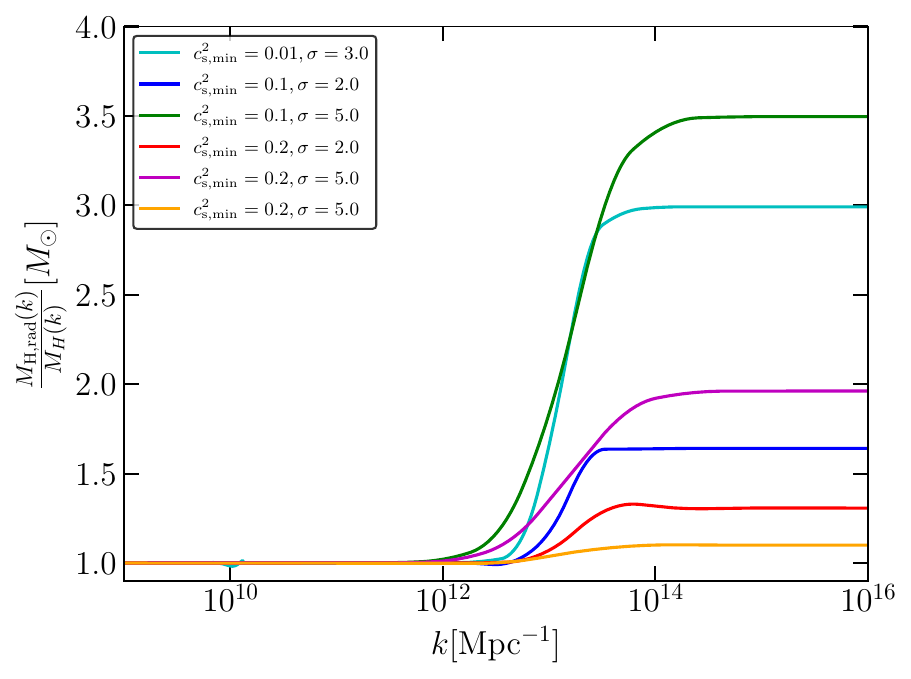}
    \includegraphics[width=2.7 in]{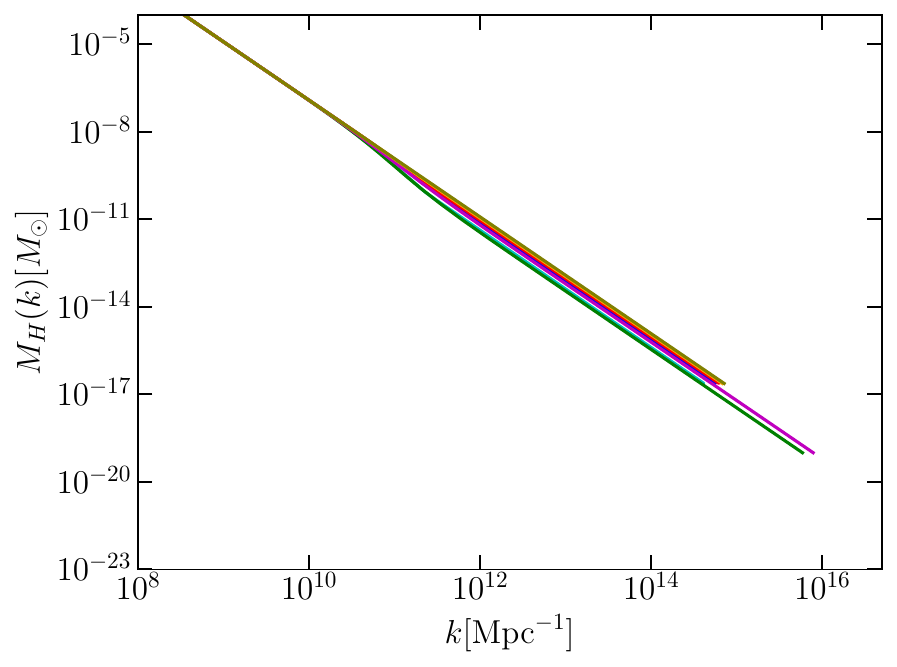}
    \includegraphics[width=2.7 in]{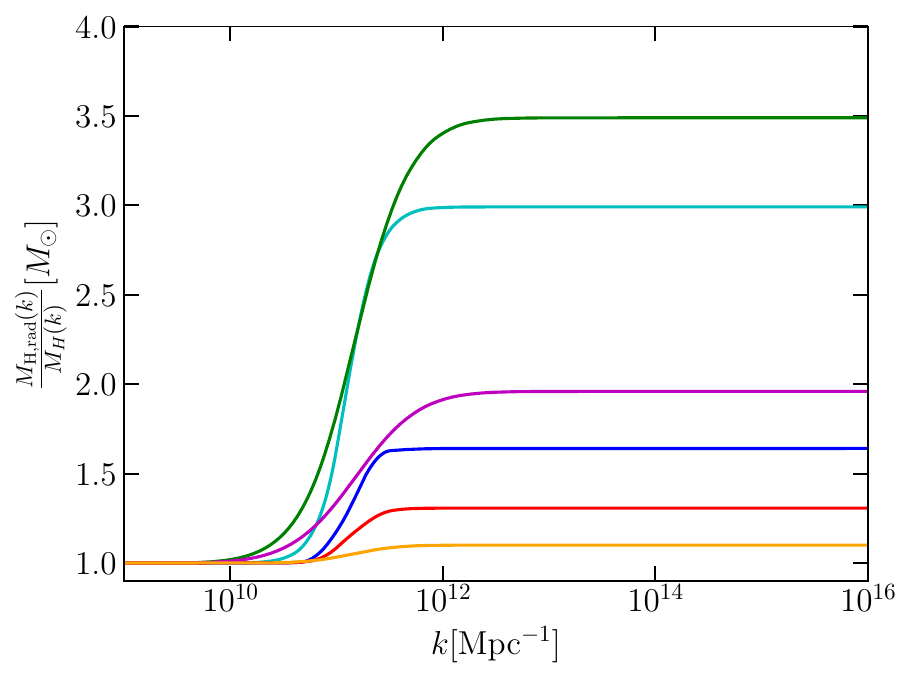}
    \caption{Mass of the cosmological horizon at horizon reentry (in solar mass) for a given wave-mode $k$ (in $\si{Mpc^{-1}}$) for different realizations of the \ac{SC}. Cases fixing $M_H(\csmin^2)=10^{-13} M_{\odot}$ (top-panels) and $M_H(\csmin^2)=10^{-9} M_{\odot}$ (bottom-panels).}
    \label{fig:horizon_mas}
\end{figure}

We numerically solve the \ac{FLRW} background equation of Eq.~(3) to obtain a relation between the mass of the cosmological horizon of given wave-mode $k$ when reenters it and itself. We fix the condition at the \ac{EW} epoch to be $M_H(t_\EW)\approx 10^{-6}M_{\odot}$ with $k_\EW \approx 10^{9}\si{Mpc^{-1}}$~\cite{Tomberg:2021ajh}, irrespective of the \ac{SC} template considered. Then the time of horizon crossing $t_H$ is given by $k =a(t_H) H(t_H) $, which can be translated to be
%Then the time of horizon crossing is given by $1 =a(t_H) H(t_H) r_\um$, which can be translated to be
%
\begin{equation}
    a^2(t_H) \rho(t_H) \tau^2 = \rho_\ub(t_0),
    \label{eq:horizon_time_crosing}
\end{equation}
for a practical numerical computation, where $\tau^{-1} = k \, R_H(t_0)$. The mass $M_{H}$ can be then evaluated at each time $t_H$ for a range of different initial conditions of $\tau$. The result is shown in Fig.~\ref{fig:horizon_mas}.

We take into account $M_{H}(k)$ when computing the induced \acp{GW} to make a fully realistic result, but not for the \ac{PBH} abundance estimation, to simplify the computation, which would have a minor impact.

%%%%%%%%%%%%%
%%%%%%%%%%%%

%\bibliographystyle{ieeetr}
\bibliographystyle{apsrev4-1}
%\nocite{*}
\bibliography{bibfile}
\end{document}